\documentclass[a4paper,11pt]{article}
\pdfoutput=1 

\usepackage{jcappub} 

\usepackage[T1]{fontenc} 
\usepackage{adjustbox}

\title{Making the leap I: Modelling the reconstructed lensing convergence PDF from cosmic shear with survey masks and systematics}

\author[a]{Alexandre Barthelemy}
\author[a,b]{Anik Halder}
\author[a,b]{Zhengyangguang Gong}
\author[c]{Cora Uhlemann}

\affiliation[a]{Universit\"{a}ts-Sternwarte, Fakult\"{a}t f\"{u}r Physik, Ludwig-Maximilians-Universit\"{a}t M\"{u}nchen,\\Scheinerstra{\ss}e 1, 81679 M\"{u}nchen, Germany}
\affiliation[b]{Max Planck Institute for Extraterrestrial Physics, Giessenbachstra{\ss}e 1, 85748 Garching, Germany}
\affiliation[c]{School of Mathematics, Statistics and Physics, Newcastle University, Herschel Building, NE1 7RU Newcastle-upon-Tyne, U.K.}

\emailAdd{A.Barthelemy@lmu.de, ahalder@usm.lmu.de, lgong@usm.lmu.de, cora.uhlemann@newcastle.ac.uk}

\abstract{The last few years have seen the development of a promising theoretical framework for statistics of the cosmic large-scale structure -- the theory of large deviations (LDT) for modelling weak-lensing one-point statistics in the mildly nonlinear regime. The goal of this series of papers is to make the leap and lay out the steps to perform an actual data analysis with this theoretical tool. Building upon the LDT framework, in this work (Paper I) we demonstrate how to accurately model the Probability Distribution Function (PDF) of a reconstructed Kaiser-Squires convergence field under a realistic mask, that of the third data release of the Dark Energy Survey (DES). We also present how weak lensing systematics and higher-order lensing corrections due to intrinsic alignments, shear biases, photo-$z$ errors and baryonic feedback can be incorporated in the modelling of the reconstructed convergence PDF. In an upcoming work (Paper II) we will then demonstrate the robustness of our modelling through simulated likelihood analyses, the final step required before applying our method to actual data.}

\usepackage{amsfonts,textcomp}
\usepackage[T1]{fontenc}
\usepackage{xcolor}
\usepackage[normalem]{ulem}

\usepackage{graphicx}	
\usepackage{amsmath}	
\usepackage{amssymb}	
\usepackage{bm}         
\usepackage{hyperref}
\usepackage{float}
\usepackage{subcaption}
\usepackage{comment}

\definecolor{color1}{rgb}{0.3, 0.7, 0}

\begin{document}
\maketitle

\section{Introduction}

In weak gravitational lensing (WL), light rays from background source galaxies propagate through the foreground inhomogeneous distribution of baryonic and dark matter. This induces (de)magnification of the brightness of galaxies and a coherent distortion pattern in their observed shapes \citep{2001PhR...340..291B, 2006glsw.conf.....M}. The statistics of the projected WL fields depend on the three-dimensional large-scale structure (LSS) and hence provide a powerful way to probe and address critical questions in cosmology such as structure formation history, the nature of dark energy and dark matter, and the laws of gravity. WL cosmology is therefore an active area of research in currently ongoing wide-area galaxy imaging surveys such as DES \citep{2016MNRAS.460.1270D}, KiDS \citep{2013ExA....35...25D} and HSC-SSP \citep{2018PASJ...70S...4A}. The promising results from these surveys have also motivated the development of new generation surveys such as Euclid \citep{Euclid} and Vera Rubin’s LSST \citep{LSST} which will soon provide data with unprecedented quality. 

One standard approach to extract cosmological information from WL fields is to analyse their power spectra (in Fourier space) or the real space 2-point correlation function (2PCF). These statistics can completely characterise the information contained in Gaussian random fields which describe our Universe accurately at its earliest stage of evolution \citep{planck} and on linear (large) scales. However, at late times, the evolution of matter density fluctuations becomes nonlinear due to gravitational instability and develops non-Gaussian features on nonlinear (small) scales. 
As a consequence, to probe the rich information contained in this non-Gaussian late-time cosmic density field one needs to continue standard 2PCF analyses as well as go beyond and investigate non-Gaussian statistical tools. We follow this path in our work.

There are various kinds of non-Gaussian statistics that have already been or that will be applied to weak-lensing survey data. We mention as examples the cosmic shear 3-point correlation function (3PCF) \cite{2003A&A...397..809S, 2005A&A...431....9S}, the integrated 3PCFs \cite{2021MNRAS.506.2780H, 2022MNRAS.515.4639H, 2023arXiv230401187G, 2023arXiv230517132H}, aperture-mass moments \cite{Barthelemy21, 2022PhRvD.105j3537S, 2023A&A...672A..44H}, lensing peaks \cite{Harnois_Deraps_2021, 2022MNRAS.511.2075Z} and density-split statistics \cite{GruenDES17, FriedrichDES17, 2023A&A...669A..69B} (to name a few). All of these have distinct advantages and disadvantages based on both their measurement and modelling strategies. A typical WL survey analysis strategy would rely on a baseline 2PCF analysis and further complement that with non-Gaussian statistics \citep{2023arXiv230112890E, 2023arXiv230401187G} to obtain stringent constraints on cosmological parameters. Extracting non-Gaussian information is a promising avenue as field-level approaches have been shown to enhance the constraining power for $\sigma_8$ and $\Omega_{\rm m}$ by a factor of 3 and 5, respectively and to break the weak lensing degeneracy \citep{Porqueres2022,Porqueres2023}. 
The Probability Distribution Function (PDF) of the weighted projection of the matter density fluctuation field along the line-of-sight, the weak lensing convergence field ($\kappa$), was shown to contain a significant amount of cosmological information \cite{2023arXiv230112890E} and to hold the potential to break cosmological parameter degeneracies that are exhibited in standard weak lensing 2PCF analyses \cite{Patton17,Liu19WLPDF, Boyle21}. By measuring the smoothed $\kappa$ field value inside apertures (or cells), the one-point $\kappa$-PDF statistic has the advantage of being straightforward to measure compared to other non-Gaussian WL convergence field probes such as bispectrum (counting triangular configurations) or Minkowski functionals (topological measurement). An emulation approach to constrain cosmological parameters using the $\kappa/\sigma$-PDF ($\sigma$ is the $\kappa$ standard deviation) jointly with convergence power spectrum was applied to HSC Year 1 (Y1) data \citep{2023arXiv230405928T}. A similar analysis was carried out for KiDS-1000 and LSST mock data as well \cite{2023MNRAS.520.1721G}. Both these works relied on small  patches obtained from N-body simulations as models for the $\kappa$-PDF to perform their cosmological analyses. In this series of work we instead use a from-first-principles theoretical modelling framework based on the large deviation theory (LDT) for the $\kappa$-PDF in the mildly nonlinear regime, giving access to the cosmological signal \citep{Boyle21} as well as the covariance \citep{Uhlemann2023cov}. This is complementary to the halo-model approach aimed to describe the more nonlinear regime \citep{2020PhRvD.102l3506M}.

However, one major difficulty in applying $\kappa$-PDF to real data analyses is that $\kappa$ itself is not a direct observable. Reference \cite{1993ApJ...404..441K} showed that the $\kappa$ and the more directly observable weak lensing cosmic shear $\gamma$ are related to each other through a convolution and pointed out that the $\kappa$ field can be reconstructed from the shear field through an inversion called the Kaiser-Squires (KS) inversion. An important point to note is that this KS reconstruction procedure is exact only if one has access to the shear field at every location on the celestial sphere (full-sky), which unfortunately is not feasible. As a consequence, since the KS inversion of the shear field is non-local, the reconstructed $\kappa$ field is severely affected by the unobserved regions on the sky due to survey geometry, the presence of masks and holes in the data as well as the field borders. The KS reconstructed $\kappa$ field can hence be significantly different from the true inaccessible $\kappa$ field. Thus, it is absolutely crucial that one includes the accurate modelling/control of the masking effect when analysing any $\kappa$-statistic from a reconstructed $\kappa$-field. To side step this problem, recent works have adopted informed priors in the reconstruction of $\kappa$ fields to infer the underlying true $\kappa$ fields, e.g.~the usage of a log-normal prior \cite{2022MNRAS.512...73F} or a sparse wavelet prior with nulled B-modes inside the survey mask \cite{2020A&A...638A.141P}. Some of these informed priors were shown to work better than the KS inversion procedure at the map-level in reconstructing the $\kappa$ field \citep{2021MNRAS.505.4626J}, and recent machine-learning-based reconstruction techniques seem to be extremely accurate \citep{2020MNRAS.492.5023J,2023A&A...672A..51R}. However, these other strategies come with the disadvantage that the impact of the reconstruction under the presence of mask is not at all straightforward to include in theoretical models of $\kappa$-field summary statistics. On the other hand, this can instead be achieved within the context of the KS reconstruction procedure and for example, has been done in the modelling of the KS reconstructed 1-point $\kappa$-map moments \cite{gattiDES} and as we shall present for the first time in this paper --- is also possible in the case of the KS reconstructed $\kappa$-PDF.

Previous works \cite{Barthelemy20a,Barthelemy21,2022arXiv221210351B} have developed an accurate theoretical model for the cumulant generating function and also the probability distribution function of the lensing-aperture mass or the $\kappa$ field using LDT\footnote{Readers interested in the application of LDT in LSS cosmology are referred to refs.~\cite{seminalLDT, saddle}.}. In this paper, we present for the first time a few-percent accurate (within cosmic variance for the DES Y3 survey area) analytical prediction of the KS reconstructed $\kappa$-PDF based on LDT while incorporating a realistic survey mask into our modelling through the pseudo-$C_{l}$ formalism \cite{Hikage2011}, mitigation of convergence $E$ to $B$ mode leakage and modification of the scaling relations between the cumulants of the density contrast field through purely geometric corrections. We show that the reconstructed $\kappa$-PDF from N-body simulations and the corresponding theoretical prediction is consistent within statistical uncertainty of cosmic variance. We also present the framework to include the modelling of different astrophysical and survey systematic effects in the reconstructed $\kappa$-PDF, including shape noise, galaxy intrinsic alignments, additive and multiplicative shear biases, photometric redshift uncertainties and higher-order lensing corrections. Using hydrodynamic simulations, we also quantify the impact of baryonic feedback on the $\kappa$-PDF and find that it mainly affects the variance of the PDF. All these components can be treated theoretically and thus significantly strengthens the case for a $\kappa$-PDF analysis in real data using a theoretical framework which is from-first-principles up to a re-scaling by the amplitude of fluctuations.

This paper is structured as follows: we first recap in section~\ref{sec::LDT} the base model for the true but observationally inaccessible (i.e.~non-reconstructed from shear field) convergence PDF from LDT. In section~\ref{sec::KS} we detail the KS reconstruction procedure of the $\kappa$ field from a simulated full-sky shear field with a realistic mask, that of the DES Year 3 (Y3) data release. We show the modification of our original LDT theoretical model of the $\kappa$-PDF to account for this reconstruction with the realistic mask in section~\ref{sec::model} and test it against measurements from N-body simulations in section~\ref{sec::test}. Furthermore, in section~\ref{sec::systematics} we present the strategy for modelling higher-order lensing, astrophysical and survey systematic effects in the $\kappa$-PDF. We summarise and conclude in section~\ref{sec::conclusion}. Appendix~\ref{appendix::masking} recaps details about the treatment of the masking effect on the cosmic shear power spectrum in the pseudo-$C_{l}$ formalism. In appendix~\ref{appendix::KS-moments} we quantify the effect of the KS reconstruction on the moments/cumulants of the reconstructed $\kappa$-field. Finally, in appendix~\ref{sec::SQUAREmask} we present a simplistic study of the impact of a square mask (without holes) on the KS reconstructed $\kappa$-PDF.

\section{Modelling the (observationally inaccessible) true convergence PDF}
\label{sec::LDT}

In this section we review how one can model the "true" but observationally inaccessible 1-point convergence $\kappa$-PDF based on large deviation theory (LDT). Readers already familiar with this model originally described in e.g.~\cite{BernardeauValageas,Barthelemy20a} can skip to section~\ref{sec::KS}.

\subsection{Statistical definitions}

We use different statistical quantities that we briefly introduce here for clarity.
From the PDF ${\mathcal P}_X$ of some continuous random variable $X$, one can define the Moment Generating Function (MGF) as the Laplace transform of the PDF
\begin{equation}
    M_X(\lambda) =\mathbb{E}\left(e^{\lambda X}\right) =  \int_{-\infty}^{+ \infty} {\rm d}x \ e^{\lambda x} {\mathcal P}_X(x) ,
    \label{laplace}
\end{equation}
or equivalently as the expectation value\footnote{Note that we make use throughout this work of the ergodicity hypothesis, in which one assumes that ensemble averages are equivalent to spatial averages ($\mathbb{E}(.)\rightarrow \left\langle.\right\rangle$) over one realisation of a random field at one fixed time. This requires that spatial correlations decay sufficiently rapidly with separation so that one has access to many statistically independent volumes within one realisation.} of the random variable $e^{\lambda X}$. Note however that the existence of an MGF is not guaranteed for all possible random fields. For example, the MGF of a strictly lognormal field is undefined for real positive $\lambda$. For the MGF to exist (in cases where the PDF also exists), the PDF needs to decay faster than the exponential $e^{\lambda x}$ for the integral in equation~\eqref{laplace} to exist. 
In the case of the one-point statistics of the cosmic density field, as computed within the large deviation framework (see section~\ref{sec::LDT_def}), there actually exists a critical positive real value $\lambda_c$ -- hereafter dubbed \textit{critical point} -- beyond which the MGF is not defined. In practice and for a field sampled within a finite volume, the MGF along the real axis will always exist and will simply tend towards $e^{\lambda X_{\rm max}}$ for $\lambda \geq \lambda_c$, where $X_{\rm max}$ is the maximum value of $X$ in the finite field.

The moment generating function, as its name implies, can be used to compute the moments of the distribution as can be seen from the series expansion of the expectation of $e^{\lambda X}$,
\begin{equation}
\begin{aligned}
    M_{X}(\lambda)\!&=\!\mathbb{E}\left(e^{\lambda X}\right)\!
    &= \sum_{n = 0}^{+ \infty} \frac{\lambda^{n} \mathbb{E}\left(X^{n}\right)}{n !},
\end{aligned}  
\end{equation}
so that the $n$th derivative of the 
MGF at $\lambda=0$ is equal to the $n$th-order moment, $\mathbb{E}\left(X^{n}\right)$.
The logarithm of the MGF is the Cumulant Generating Function (CGF) 
\begin{equation}
    \phi(\lambda) = \log(M(\lambda)) = \log(\mathbb{E}(e^{\lambda X})) = \sum_{n=1}^{+ \infty} k_{n} \frac{\lambda^{n}}{n !}\,,
    \label{eq:cgf_def} 
\end{equation}
where $k_n$ are the cumulants (i.e.~the connected moments) of the distribution and where we have dropped the subscript $X$ for clarity.

The reduced cumulants are defined as 
\begin{equation}
    S_n = \frac{k_n}{k_2^{n-1}}, \ n\geq 1
    \label{Sp}
\end{equation}
where $k_2$ is the variance. These are important in the context of cosmological structure formation because the $S_n$ of the cosmic matter density field have been shown to be independent of redshift down to mildly nonlinear scales \citep{Peebles,1995MNRAS.274.1049B} and thus introduce relevant scaling relations between the cumulants. We thus also define the scaled cumulant generating function (SCGF hereafter) as
\begin{equation}
    \varphi(\lambda) = \lim_{k_2 \rightarrow 0} \sum_{n=1}^{+\infty} S_n \, \frac{\lambda^n}{n!}=\lim_{k_2 \rightarrow 0} k_2 \,\phi\left(\frac{\lambda}{k_2}\right),
    \label{defscgf}
\end{equation}
which we also extrapolate to non-zero values of the variance (i.e.~evaluate at finite $k_2$ on our chosen smoothing scale).
One can reconstruct the PDF from the CGF using an inverse Laplace transform (inverting equation~\eqref{laplace}) 

\begin{equation}
    {\mathcal P}(x) = \int_{-i\infty}^{+i\infty} \frac{{\rm d}\lambda}{2\pi i} \, \text{exp}\left(-\lambda x + \phi(\lambda)\right)\,.
    \label{eq:inverse_laplace}
\end{equation}

\subsection{Large deviation theory of the matter density field}
\label{sec::LDT_def}

The large deviation theory (LDT) framework in large-scale structure has mainly been used to model the one-point statistics of the smoothed 3D matter density field \citep[see, for example,][]{seminalLDT, saddle, cylindres} and also of projected 2D quantities such as weak lensing convergence and aperture-mass fields \citep{Barthelemy20a,Barthelemy21,Boyle21}.
It has also been extended to the joint distribution between densities measured at some distance \citep{2016MNRAS.460.1598C} and projected quantities between different source redshift bins \citep{Barthelemy22} (i.e.~n-point PDF). The results are most simply applied for highly symmetrical window functions such as two- or three-dimensional top-hats, but can be generalised to other smoothing schemes \citep{seminalLDT, paolo, Barthelemy21}. We begin here by recalling some of the results of LDT for the one-point statistics of the matter density contrast smoothed in two-dimensional disks (which replicates the dynamics within long cylinders), which in turn will allow us to compute the one-point statistics of projected 2D quantities like the convergence field.

A set of random variables $\{\rho^\epsilon\}_{\epsilon}$ with PDF ${\mathcal P}_{\epsilon}(\rho^\epsilon)$ is said to satisfy a large deviation principle if the limit
\begin{equation}
    \Psi_{\rho^\epsilon}(\rho^\epsilon) = - \lim_{\epsilon \rightarrow 0} \epsilon \log\left[{\mathcal P}_{\epsilon}(\rho^\epsilon)\right]
    \label{LDP}
\end{equation}
exists, where $\epsilon$ is the \textit{driving parameter}. $\Psi$ is known as the rate function of $\rho^\epsilon$ and describes the exponential decay of its PDF. The driving parameter $\epsilon$ indexes the random variables with respect to some evolution, for example an evolution in time.
In the case of the matter density field smoothed on a single scale $R$, this driving parameter is the variance, which acts as a clock from initial to late times ($\epsilon \equiv \sigma^2_R$). We now omit the $\epsilon$ sub/superscripts in our notation for simplicity.

The existence of a large deviation principle for the random variable $\rho$ implies that its SCGF $\varphi_{\rho}$ is given through Varadhan's theorem as the Legendre-Fenchel transform of the rate function $\Psi_{\rho}$ \citep{Ellis_book,2009PhR...478....1T,touchette}
\begin{equation}
    \varphi_{\rho}(\lambda) = \sup_{\rho} \,\left[ \lambda\rho - \Psi_{\rho}(\rho)\right],
    \label{varadhan}
\end{equation}
where the Legendre-Fenchel transform reduces to a simple Legendre transform when $\Psi_{\rho}$ is convex. In that case, 
\begin{equation}
    \varphi_{\rho}(\lambda) =  \lambda\rho - \Psi_{\rho}(\rho),
    \label{Legendre}
\end{equation}
where $\rho$ is a function of $\lambda$ through the stationary condition\footnote{The $\rho_c$ value at which $\Psi_{\rho}$ ceases to be convex leads to a $\lambda_c$ value which corresponds to the critical point mentioned when discussing equation~\eqref{laplace}.} 
\begin{equation}
   \lambda = \frac{\partial \Psi_{\rho}(\rho)}{\partial \rho}.
    \label{stationnary}
\end{equation}
Another consequence of the large-deviation principle is the so-called contraction principle.
This principle states that for a random variable $\tau$ satisfying a large deviation principle and related to $\rho$ through the continuous mapping $f$, the rate function of $\rho$ can be computed as
\begin{equation}
    \Psi_{\rho}(\rho) = \inf_{\tau:f(\tau) = \rho} \Psi_{\tau}(\tau).
    \label{contraction}
\end{equation}
This is called the contraction principle because $f$ can be many-to-one, in which case we are {\it contracting} information about the rate function of one random variable down to the other. In physical terms, this states that an improbable fluctuation of $\rho$ is brought about by the most probable of all improbable fluctuations of $\tau$. 

For the case of the normalised 3D matter density field $\rho \equiv \rho/\bar{\rho}$, the rate function of the late-time normalised density field at different scales can be computed from initial conditions if the most likely mapping between the two is known -- that is, if one is able to identify the leading field configuration that will contribute to the infimum of equation~\eqref{contraction}. In cylindrically symmetric configurations, as for a disk of radius $R$ at redshift $z$ (in 2D space) or alternatively a very long 3D cylinder centered on this disk,
the most likely mapping between final and initial conditions should preserve the symmetry \citep{Valageas,2019JCAP...03..009I}\footnote{This is only true for a certain range of density contrasts around zero, very much sufficient for our purposes. However, one could note that there are counter-examples in which the spherical or cylindrical symmetry does not lead to spherical/cylindrical collapse being the most likely dynamics, for example in 1D for very high values of the density.}. This in turn leads to initial conditions also being cylindrically symmetric and the dynamics between the two being that of cylindrical collapse. 

Thus, starting from Gaussian initial conditions\footnote{Primordial non-Gaussianity can also straightforwardly be accounted for in this formalism as shown by \cite{NonGaussianities,Friedrich_2020_pNG}.}, the rate function $\Psi_{\bar{\tau}}$ of the most probable initial (linear) density field $\bar{\tau}$ is simply a quadratic term. From the dynamics of cylindrical collapse that maps the most probable initial and late-time fields, the rate function $\Psi_{\rm cyl}$ of the normalised late-time density field $\rho$ in a disk of radius $R$ is then given by
\begin{equation}
    \Psi_{\rm cyl}(\rho)=\sigma^2_{R}\frac{\bar{\tau}^2}{2 \sigma^2_{r}},
    \label{psicyl}
\end{equation}
where $\sigma^2_{R}$ --- our driving parameter --- is the variance of the nonlinear density field in the disk, $\sigma^2_{r}$ is the variance of the linear density field inside the initial disk (before collapse) of radius $ r = R \, \rho^{1/2}$ (from mass conservation), and $\bar{\tau}$ is the linear density contrast obtained through the most probable mapping between the linear and late-time density fields.
This mapping is given by 2D spherical (cylindrical) collapse, for which an accurate parametrisation is given by \citep{Bernardeau1995}
\begin{equation}
    \zeta(\bar{\tau}) = \rho =  \left(1 - \frac{\bar{\tau}}{\nu} \right)^{-\nu}.
    \label{collapse}
\end{equation}
This parametrisation is in the spirit of previous works involving the density filtered in spherical cells, while the value of $\nu$ in this parametrisation of $\zeta$ is chosen to be $\nu = 1.4$, so as to reproduce the value of the leading-order (tree-level) skewness in cylinders as computed from Eulerian perturbation theory \citep{cylindres}.

Then, as a straightforward consequence of the contraction principle, the rate function given by equation~\eqref{psicyl} is also the rate function of any monotonic transformation of $\rho$, so that for the density contrast $\delta = \rho - 1$, we have $\Psi_{\delta}(\delta) = \Psi_{\rho}(\rho(\delta))$.
Thus, plugging equation~\eqref{psicyl} into equation~\eqref{Legendre} gives us the SCGF $\varphi_{\rm cyl}$ of the matter density contrast in a disk at redshift $z$. 

Finally, one of the key aspects of the large deviation formalism in the cosmological context is that we apply the result for the SCGF beyond the $\sigma^2_R \rightarrow 0$ limit. This extrapolation of the exact result allows us to obtain a realistic CGF $\phi_{\rm cyl}$ of the real density field for non-vanishing $\sigma^2_R$ by rescaling the SCGF by the driving parameter (the nonlinear variance) at the scale and redshift being considered. This yields:
\begin{equation}
    \phi_{\rm cyl}(\lambda) = \frac{1}{\sigma^2_R}\varphi_{\rm cyl}(\lambda \sigma^2_R). 
\end{equation}
This is physically meaningful because we thus construct a CGF naturally matching its quasi-linear limit and since the reduced cumulants $S_n$ from the cylindrical/spherical collapse dynamics have been shown to be very robust over a large range of scales and redshifts down to mildly nonlinear scales ($\gtrsim 5$ Mpc/$h$ at $z \gtrsim 0$, see for example figure~2 in \cite{saddle} and figure~A1 in \cite{cylindres}) so that rescaling by the nonlinear variance $\sigma^2_{R}$ allows access to the full one-point statistics of the nonlinear density field. In LDT terms, the SCGF given by the large deviation principle is the well-defined asymptotic form taken by the reduced cumulants of the field in the regime where the variance goes to zero, 
and we simply keep this form for our predictions of the nonlinear CGF. 

Finally, note that though equations~\eqref{Legendre} and \eqref{psicyl} have been known and used for three decades in the context of count-in-cells statistics, that is the density field filtered in 3D top-hat windows \citep[see for example][]{Bernardeau_1994a,Valageas_1998}, their re-derivation through large deviation statistics is more general and allows to set up a framework for the computation of different probabilities in the cosmological context.

\subsection{From matter density to convergence}

Let us recall that for a flat cosmology, the convergence $\kappa$ can be interpreted as a line-of-sight projection of the matter density contrast between the observer and the source and can be written as \citep{kappadef}
\begin{equation}
    \kappa({\bm \vartheta}) = \int_0^{+\infty} {\rm d}z \frac{{\rm d}\chi}{{\rm d}z} \, \omega_{n(z)}(\chi) \, \delta(\chi,\chi{\bm \vartheta}),
    \label{def-convergence}     
\end{equation}
where $\chi$ is the comoving radial distance and the generalised lensing kernel $\omega_{n(z)}$ for a wide distribution of sources following the normalised distribution $n(z)$ is
\begin{equation}
    \label{eq:lensing_kernel}
    \omega_{n(z)}(z) = \frac{3\Omega_{\rm m} H_0^2}{2 c^2} \int \!{\rm d}z_\mathrm{s} \, n(z_\mathrm{s})  \frac{[\chi(z_\mathrm{s})-\chi(z)]\,\chi(z)}{\chi(z_\mathrm{s})} \mathcal{H}(z_\mathrm{s}-z) (1+z),
\end{equation}
where the Heaviside $\mathcal{H}$ ensures that the integrand vanishes for $z \geq z_\mathrm{s}$. Hereafter we only note the lensing kernel $\omega$ for simplicity.
Under the small-angle/Limber approximation, it can be shown that correlators of the (smoothed) convergence field can be seen as a juxtaposition (by which we mean an integral along the line-of-sight) of the 2D correlators of the underlying density field, as if each 2D slice along the line-of-sight is statistically independent of the others \citep{BernardeauValageas,Barthelemy20a}. In terms of the one-point statistics of the smoothed $\kappa$ field within a top-hat window function of angular radius $\theta$, this translates to saying that the CGF of $\kappa$ is a sum along the line-of-sight of the CGF of independent 2D slices of the matter density contrast:\footnote{Rigorously, this result applies for a juxtaposition of very long cylinders centered on the slices and of length $L\rightarrow
\infty$. Since the symmetry and thus the most likely dynamics of these long cylinders are the same as for a 2D slice in a 2D space, and since the results are independent of $L$, we refer to \lq 2D slices\rq~for clarity. This emphasises that correlations along the line-of-sight are negligible compared to those in the transverse directions.}
\begin{equation}
    \phi_{\kappa,\theta}(\lambda) = \int_0^{+\infty} {\rm d}z \frac{{\rm d}\chi}{{\rm d}z} \, \phi_{{\rm cyl}}^{<\chi(z) \theta}(\omega_{n(z)}(z) \lambda,\chi(z)),
    \label{projection}
\end{equation}
where $\phi_{{\rm cyl}}^{<\chi(z) \theta}$ is the CGF of the density contrast filtered within a disk of radius $\chi(z) \theta$ so as to reproduce the geometry of the light cone.

Equation~\eqref{projection} thus reduces the complexity of the problem down to computing the one-point statistics of the 2D matter density in each two-dimensional slice (or equivalently within long 3D cylinders at the same redshift up to some factor depending only on the length of the cylinder)
along the line-of-sight, which we have already done in section~\ref{sec::LDT_def}. Using these results, we can then build the nonlinear CGF $\phi_{\kappa,\theta}$ of the convergence field. Note that equation~\eqref{projection} highlights the nice property of the projected CGF being expressible simply as a sum of independent redshift slices. This is an important property when considering more complicated joint distributions (e.g. joint convergence CGF of multiple source tomographic redshift bins), in which the only modification in this integral would be the replacement of the $\omega_{n(z)}(z)\lambda$ term by a term depending on more than one $\lambda$ variable. This form is much simpler than that of the corresponding multi-variate PDF \citep{Barthelemy22}.

Finally, it is important when working with the CGF (albeit to compute the PDF) to know the approximate location of the (theoretical) critical point of the convergence field, $\lambda_c$. First, the critical points $\delta_c$ in each redshift slice are calculated by finding where the second derivative of the rate function becomes zero. The corresponding critical $\lambda$ values can be obtained by applying the stationary condition (equation~\eqref{stationnary}). The minimum $\lambda_c$ along the line-of-sight is then taken as the critical point of the convergence CGF.

The convergence PDF $\mathcal{P}(\kappa)$ is then computed from its CGF $\phi_{\kappa,\theta}$ using the inverse Laplace transform of equation~\eqref{eq:inverse_laplace} as 
\begin{equation}
    \mathcal{P}(\kappa) = \int_{-i\infty}^{i\infty} \frac{{\rm d} \lambda}{2 \pi i} \exp\left(-\lambda \kappa + \phi_{\kappa, \theta}(\lambda)\right),
\end{equation}
and thus explicitly depends on the lensing kernel $\omega_{n(z)}$ from equation \eqref{eq:lensing_kernel}.
This computation involves a continuation of the CGF in the complex plane that can be performed using different techniques. We refer to \cite{Barthelemy20a} and \cite{Barthelemy21} for technical discussions on the two possible methods that have been used in the literature to perform this complex continuation in the LDT context. Here, we use the one of \cite{Barthelemy21} that relies on an informed fitting function of the CGF.

\section{Measuring the KS reconstructed convergence PDF}
\label{sec::KS}

In this section we describe how one performs a Kaiser-Squires inversion on an observed masked shear field in order to create a reconstructed convergence $\kappa$ map as well as how one can measure a meaningful $\kappa$-PDF from this map.

\subsection{From galaxy ellipticities to shear field}

In real-life weak-lensing experiments, the convergence field is not a direct observable (except maybe through magnification, see for example \cite{2015MNRAS.452.1202A}). What we actually observe are the shapes of source galaxies, their ellipticities, which are a noisy estimate of the reduced shear $g$:
\begin{equation}
    \epsilon = \frac{g + \epsilon_{\rm IA} + \epsilon_{\rm n}}{1+g(\epsilon_{\rm IA} + \epsilon_{\rm n})}
\end{equation}
with $g = \gamma/(1-\kappa)$, $\epsilon_{\rm IA}$ the intrinsic shape of the galaxy and $\epsilon_{\rm n}$ the shape measurement noise. In the weak-lensing regime, both the shear $\gamma$ and the convergence $\kappa$ are $\ll 1$ so that the ellipticities can serve as a noisy estimate of the shear field through
\begin{equation}
\label{eq:observed_ellipticity}
    \epsilon \approx \gamma + \epsilon_{\rm IA} + \epsilon_{\rm n}.
\end{equation}

In the following we will ignore intrinsic alignments whose introduction in the modelling will be described later in section~\ref{section::IA}. On the other hand, the contribution of noise is estimated in the literature by randomly rotating the shape of the galaxies to erase the cosmological contribution which would lead to pure-noise ellipticity, shear and then convergence fields (whose reconstruction we detail below) of zero average but non-negligible variance. By virtue of the central limit theorem, this noise is expected to become Gaussian for large numbers of galaxies and is also expected to be closely independent from the reconstructed field. Since the simulations we use in this work contain no intrinsic shape noise, we will assume that those two hypotheses are valid and shape noise can thus be trivially taken into account in the theoretical modelling of the convergence PDF by simple convolution of the clean PDF with a Gaussian of the correct shape noise amplitude (we will test this in Paper II of the series). For a survey with varying source galaxy density across the sky, this would lead to a mean shape noise variance which would still convolve the clean PDF. Note however that the Gaussian hypothesis can be easily lifted if needed -- simply by replacing the Gaussian convolution by another one -- as well as the hypothesis of zero correlation between the noise and reconstructed field which can be tested by computing the joint PDF between the noise map (obtained through random rotations) and the observed convergence. A denoised estimator of the observed convergence would then be obtained by marginalisation over the noise in this joint PDF.

\subsection{From shear to convergence}
\label{sec::gamma2kappa}

Crucially, for a footprint of approximately 5000 deg$^2$ such as the one from the DES Y3 -- and up to three times that for a Euclid-like experiment -- we need a full-sky, spherical harmonics approach to estimate the convergence field from the shear \citep{2021MNRAS.505.4626J, 2022MNRAS.509.4480W}. In this formalism, the (Born-)projected Poisson equation reads \citep{2005PhRvD..72b3516C}
\begin{equation}
    \kappa = \frac{1}{4}(\eth\bar{\eth}+\bar{\eth}\eth) \psi,
\end{equation}
with $\psi$ the usual projected (under Born approximation) gravitational potential and $\eth$, $\bar{\eth}$ are respectively the raising and lowering operators acting on the spin-weighted spherical harmonics \citep[see appendix~A of][]{2005PhRvD..72b3516C}. Similarly, the usual complex shear field equations translate into
\begin{equation}
    \gamma = \gamma_1 + i\gamma_2 = \frac{1}{2} \eth\eth \psi.
\end{equation}

Finally, expanding the projected gravitational potential and convergence (scalar, spin-0 fields) as well as the spin-2 complex shear on the spin-weighted spherical harmonics basis $_0\!Y_{lm}$ and $_2\!Y_{lm}$ respectively, we get
\begin{align}
    &\psi(\theta,\phi) = \sum_{l,m} \psi_{lm}\,  _0\!Y_{lm}(\theta, \phi) , \\
    &\kappa = \kappa_E + i \kappa_B = \sum_{l,m} (\kappa_{E,lm}+ i \kappa_{B,lm}) \, _0\!Y_{lm} , \\
    &\gamma = \gamma_1 + i \gamma_2 = 2 \sum_{l,m} (\gamma_{E,lm}+ i \gamma_{B,lm})\,  _2\!Y_{lm}.
\end{align}

Both the convergence and the shear have been decomposed into curl-free E-modes and divergence-free B-modes, and we can relate the shear to the convergence as
\begin{align}
    &\kappa_{E,lm}+ i \kappa_{B,lm} = \frac{1}{2} l(l+1)\psi_{lm} , \\
    &\gamma_{E,lm}+ i \gamma_{B,lm} = \frac{1}{2} \left[l(l+1)(l-1)(l+2)\right]^{1/2}\psi_{lm} , \\
    \implies & \kappa_{E,lm}+ i \kappa_{B,lm} = \sqrt{\frac{l(l+1)}{(l+2)(l - 1)}} (\gamma_{E,lm}+ i \gamma_{B,lm}).
    \label{KS_sphere}
\end{align}
Equation \eqref{KS_sphere} is the harmonic space spherical-sky generalisation of the Kaiser-Squires (KS) inversion formula (originally proposed by \cite{1993ApJ...404..441K} for flat-sky convergence field reconstruction from the observed shear field). 
An inverse spherical-harmonic transform on the full-sky allows one to then obtain the KS reconstructed $\kappa_E$ and $\kappa_B$ fields on the celestial sphere.
Strictly speaking, the equations above are correct assuming no couplings between the lenses that would appear when relaxing the Born approximation and taking into account corrections of the same order in the Sachs' equation of gravitational lensing \citep{Sachs61,Barthelemy20b}. In this paradigm, the only allowed B-modes are those produced by the effect of masking the full-sky or observational systematics. On the other hand, allowing for couplings between lenses along the line-of-sight would slightly change the definitions of the shear and convergence and as a consequence their relationship. This is not a fundamental issue since those higher-order corrections can be modelled respectively for the shear and the convergence in cases where one or the other can be accessed \citep[see for example][]{Fabbian18,Fabbian19,Barthelemy20b}, and since these corrections are small, one could still reconstruct a field from equation~\eqref{KS_sphere} and call it \textit{convergence}. Alternatively, the knowledge of what are both the true shear and convergence has also permitted to build Bayesian reconstruction schemes of the convergence based on an observed shear field with the help of informed priors \citep[see for example][]{2020A&A...638A.141P,2021MNRAS.505.4626J,2022MNRAS.512...73F} as well as machine learning reconstruction algorithms \citep{2020MNRAS.492.5023J} and more recently clever mix of the two \citep{2023A&A...672A..51R}. In this paper, we reconstruct a convergence field by applying the KS inversion equation~\eqref{KS_sphere} on a masked full-sky shear field which contains higher-order lensing corrections and thus admits B-modes. We then only work with the reconstructed $\kappa_E$ field and theoretically model the E-modes one-point PDF assuming that the main difference between the reconstructed and "true" PDFs are sourced only by masking the full-sky shear field\footnote{The theoretical model for the non-reconstructed $\kappa$-PDF does not natively include higher-order lensing corrections (though they can also be included in the model if desired as discussed in section~\ref{sec::higher-order-corrections}) hence the quotation marks around "true". However, the simulated convergence fields we use are either obtained through ray-tracing hence beyond Born-approximation or reconstructed from ray-traced shear fields thus admitting physical B-modes. We have thus checked explicitly that those are fortunately negligible at the scales and redshifts we consider.}.

\subsection{Reconstructing the convergence PDF from simulated shear maps}
\label{sec::reconstruct_PDF}
\begin{figure}
    \centering
    \includegraphics[width = 0.9\columnwidth]{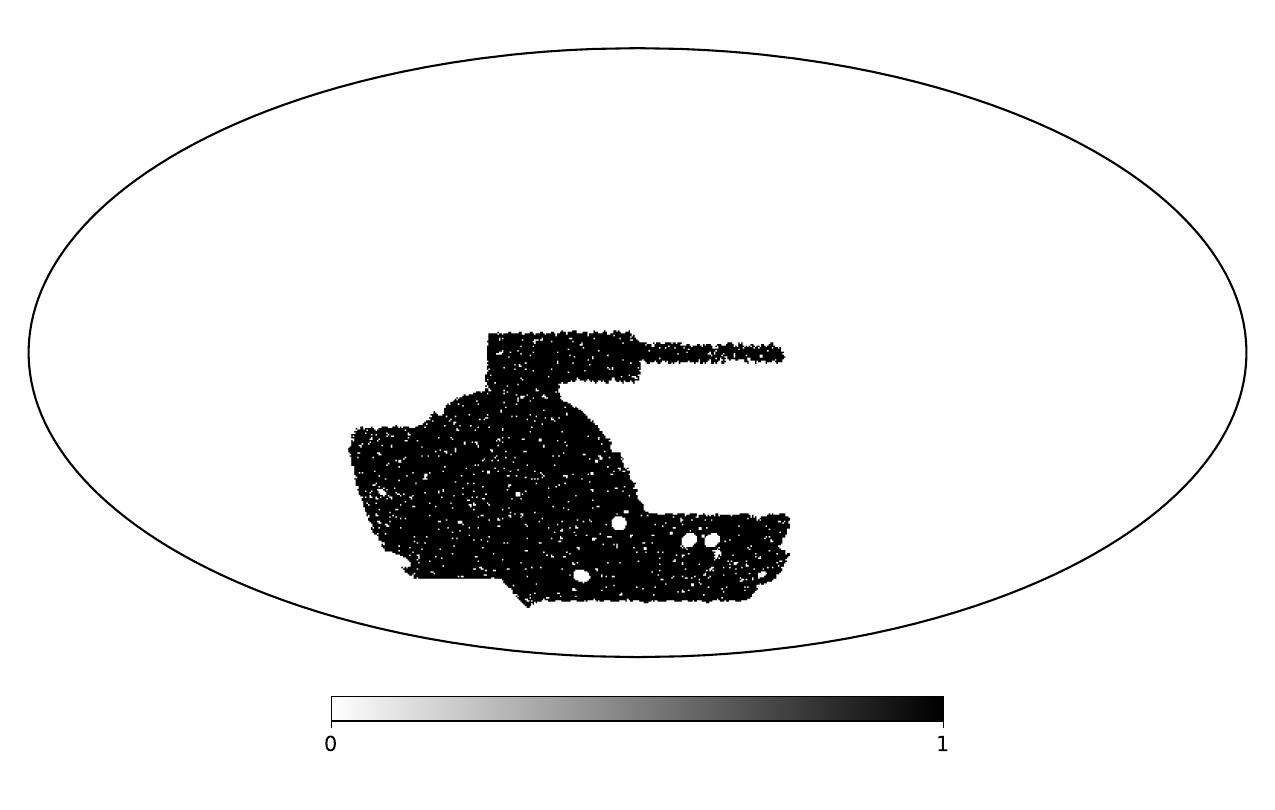}
    \caption{Illustrative Mollweide projection of the binary mask applied to the full-sky shear field to mimic DES Y3 observations. The mask is as close as possible to the real one, notably keeping the multiple holes of different sizes across the field of view. The observed fraction of the full-sky is $f_{\rm sky} = 0.1149$.}
    \label{fig:mask}
\end{figure}

\begin{figure}
    \centering
    \includegraphics[width = 0.9\columnwidth]{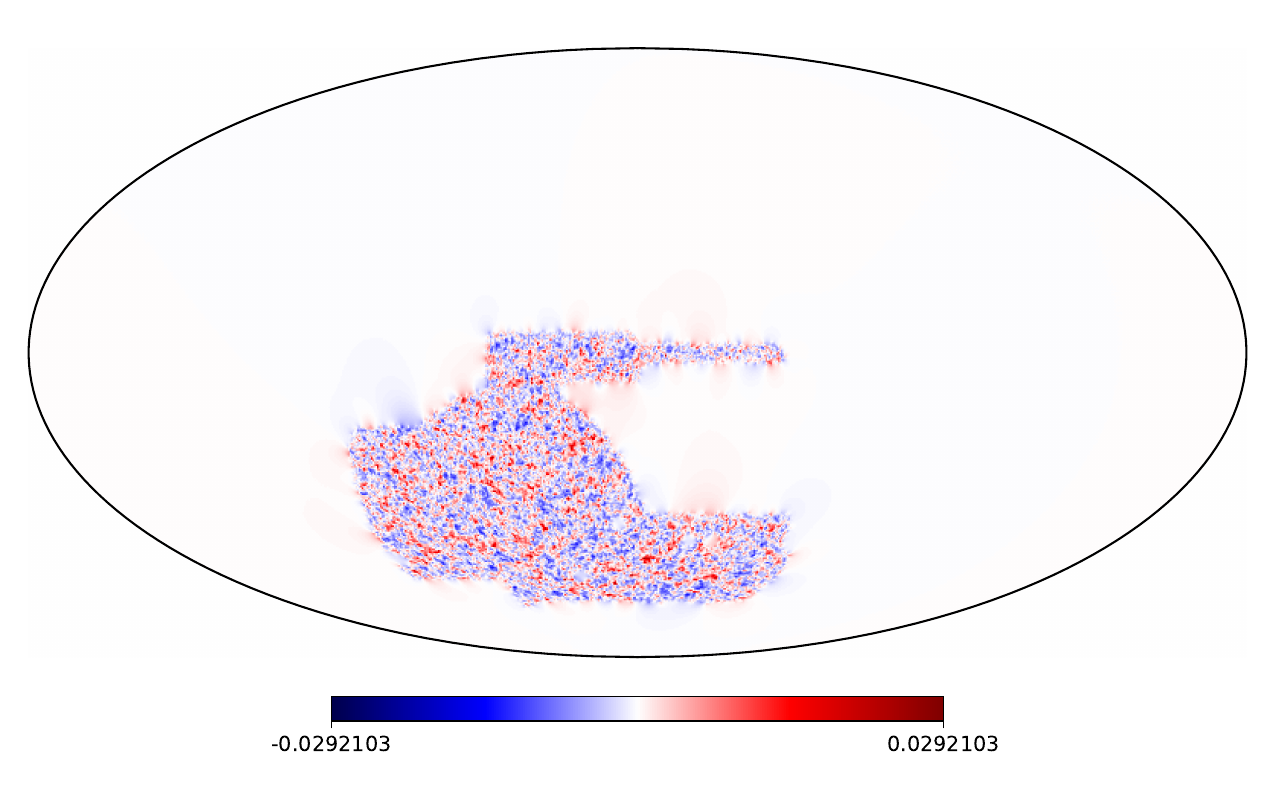}
    \caption{Illustrative Mollweide projection of the full-sky reconstructed $\kappa_{E,\rm FS}$ field for sources mimicking the DES Y3 fourth redshift bin and from a simulated shear field under the DES Y3 mask shown in figure~\ref{fig:mask}. The reconstructed $\kappa_E$ field has been further smoothed by a top-hat window of radius 20 arcmin. Observed carefully, one can spot the non-locality of the KS reconstruction through the blurring of the mask boundaries and holes. It should also be noted that the majority of the pixel values far outside the survey footprint fluctuate very closely around zero (but are not exactly zero), as expected.}
    \label{fig:kappa_E}
\end{figure}

In actual observations, one would start from a discrete shear catalogue that has to be interpolated (pixelised) to create a shear map. We here skip that step and directly use a set of publicly available full-sky weak lensing shear maps in \verb|Healpix| format \citep{2005ApJ...622..759G}\footnote{\url{https://healpix.sourceforge.io/}} generated from the ray-traced N-body simulation suite from Takahashi et al. \citep{takahashi_full-sky_2017}\footnote{\url{http://cosmo.phys.hirosaki-u.ac.jp/takahasi/allsky_raytracing/}}. We have 108 realisations of several fixed source redshifts up to $z_s \sim 5$ and combine them to mimic a realistic source distribution from an input $n(z)$ (see section~\ref{sec::test} for the description of how we do that in practice to mimic the 4$^{\rm th}$ source redshift bin of the DES Y3).
For each set of full-sky shear maps, we multiply those by a binary mask, shown in figure~\ref{fig:mask}, mocking observed maps from DES Y3 and reconstruct the E-modes full-sky convergence field using KS reconstruction equation~\eqref{KS_sphere}. This reconstruction makes use of the functions {\sc map2alm} and {\sc alm2map} of the \verb |Healpy| package \citep{Zonca2019}. We then smooth the maps with top-hat filters of radius $\theta$ in harmonic space and the only remaining step is to select the pixels on the full-sky map to then construct a meaningful/physical convergence PDF. Indeed, since large parts of the shear field were set to zero due to the binary survey mask (and though the reconstruction of the convergence is in principle non-local) the reconstruction of $\kappa_E$ in the previously masked areas yields unphysically small values compared to the fluctuations of the true convergence $\kappa_{\rm true}$ values (obtained directly from the simulation suite) in those regions. We illustrate this point in figure~\ref{fig:kappa_E} where we display a full-sky (FS) reconstructed field $\kappa_{E, \rm FS}$  smoothed with a top-hat of radius 20 arcmin.
We thus only take into account pixels whose smoothed convergence values $\kappa_{E,\rm unmasked}$ results from unmasked regions on the sky\footnote{Measuring the full-sky convergence PDF (i.e.~using all the pixels over the entire sky) reconstructed from a partial sky shear field would yield a distribution with very high kurtosis (highly peaked around zero) illustrating the fact that most values are close to zero around which we observe fluctuations that correspond to the unphysically small (close to zero) reconstructed convergence values in the masked regions.}. That way, we can expect a loss of power of $\kappa$ E-modes to B-modes in the reconstruction due to the presence of the mask and the non-locality of the transform, but sufficiently mitigated such that physical scaling relations for example between higher-order cumulants of the matter density field still hold. We illustrate this fact more quantitatively in the next section in equation~\eqref{eq::schematic_sigma}. In practice, this can be done by smoothing the binary mask with the same top-hat kernel as the field, and only keeping pixels where the smoothed masked has values sufficiently close to 1. For the DES Y3 mask, we find in practice that keeping all pixels where the smoothed mask is higher than 0.98 is a good equilibrium between keeping the maximum possible portion of the survey volume and mitigating the influence of masked pixels.
This approach might be improved by upweighting the pixels that are partially masked as done in the density-split statistics context \cite{FriedrichDES17,GruenDES17} which is similar to ours. It might allow to consider more pixels but we do not implement here such a strategy.

\section{Modelling the KS reconstructed convergence PDF}
\label{sec::model}

Having described the LDT based framework for modelling the observationally inaccessible $\kappa$-PDF in section~\ref{sec::LDT} and then the actual procedure to reconstruct and measure the $\kappa$-PDF under the presence of a realistic survey mask using KS inversion in section~\ref{sec::KS}, we now describe how to modify our original framework to incorporate the effect of the survey mask and develop a model for the KS reconstructed $\kappa$-PDF.

\subsection{Schematic view}

Let us first have a look at how the reconstruction scheme presented in the previous section affects the amplitude of convergence fluctuations. Schematically, masking the field reduces the overall shear power spectrum and thus the reconstructed convergence power spectrum by a factor $f_{\rm mix}$\footnote{Note that this is an effective and schematic treatment. In reality the loss of power is modelled by the convolution of the underlying true power spectrum by a mode-mixing matrix as described in appendix~\ref{appendix::masking}.} due to mode-mixing which is at first order comparable to the observable fraction of the full-sky $f_{\rm sky}$. On the other hand, the reconstructed convergence in the masked regions (e.g. far away from the survey footprint) is mostly 0 so that we could model the full-sky reconstructed PDF as a sum of the distribution of $\kappa_E$ values in the unmasked regions and a Dirac-delta distribution, both weighted by $f_{\rm sky}$ and $(1-f_{\rm sky})$ respectively. As a consequence, the variance of the reconstructed $\kappa_E$ in the unmasked region of the sky $\sigma^2_{\kappa_E, {\rm unmasked}}$ is roughly the variance $\sigma^2_{\kappa_E, {\rm FS}}$ of the full-sky $\kappa_{E,\rm FS}$ reconstructed from the masked shear field divided by $f_{\rm sky}$ and we have 
\begin{equation}
    \sigma^2_{\kappa_E, {\rm unmasked}} \approx \frac{1}{f_{\rm sky}}\sigma^2_{\kappa_E, {\rm FS}} \approx \underbrace{\frac{f_{\rm mix}}{f_{\rm sky}}}_{O(1)}\sigma^2_{\kappa, {\rm true}} \ .
    \label{eq::schematic_sigma}
\end{equation}
This is further illustrated with actual values of the $\kappa_E$ variance under the DES Y3 mask in appendix~\ref{appendix::KS-moments}.

Additionally, the core idea behind the large deviation approach to the statistics of the cosmic matter density field is to identify asymptotic scaling relations between the cumulants of the field such that specifying the value of the variance serves both as a dial controlling the proximity to the asymptotic limit and as a closure relation allowing to compute all cumulants of the field from the variance. For now neglecting the projection along the line-of-sight (and since the convergence reconstruction scheme only affects the amplitude of fluctuation mildly), it could seem natural to assume that at first order the reconstruction affects the successive cumulants of the convergence field in a manner consistent with their scaling relations with the variance based on constant reduced cumulants $S_n$ (see equation \eqref{Sp}). This would typically mean that the loss of power from E to B modes in the region of the sky where it is mitigated the most (which we call the "unmasked" regions, see section~\ref{sec::reconstruct_PDF}), preserves core physical properties of the field. As such, neglecting the projection along the line-of-sight, the $n^{\rm th}$ cumulants of the reconstructed $\kappa_E$ field in the unmasked regions can schematically be related to the cumulants of the true but observationally inaccessible $\kappa$ field as
\begin{equation}
    \langle \kappa_E^n \rangle_{c, {\rm unmasked}} \approx \left(\frac{f_{\rm mix}}{f_{\rm sky}}\right)^{n-1} \langle \kappa^n \rangle_{c, {\rm true}}.
    \label{eq::schematic_k3}
\end{equation}
In the above equation~\eqref{eq::schematic_k3}, the subscript $c$ denotes the connected parts of the moments, i.e.~the cumulants.

\subsection{Reconstructed cumulant generating function}

More precisely, the large deviation principle applied to the cosmic matter density field implies that the scaling relations (constant $S_n$ in equation \eqref{Sp}) between cumulants are correct for the matter density field, but not for the projected convergence field itself. However, under the Born and Limber approximations, the total lensing effect that we observe can be treated as the independent combination of successive lensing events along the line-of-sight. As a consequence, taking into account the convergence reconstructing scheme while preserving the scaling relations between cumulants must be understood at the level of each lensing event, that is for the cumulants of the matter density contrast along the line-of-sight. In terms of the reconstructed CGF $\phi_{\kappa_E}$, this can be written by modifying equation~\eqref{projection} as
\begin{equation}
    \phi_{\kappa_E, \theta}(\lambda) = \int_0^{+\infty} {\rm d}z \frac{{\rm d}\chi}{{\rm d}z} \, \frac{\langle \delta^2_{\rm true} \rangle}{\langle \delta^2_E \rangle} \phi_{{\rm cyl}}^{<\chi \theta}\left( \frac{\langle \delta^2_E \rangle}{\langle \delta^2_{\rm true} \rangle}\omega(\chi) \lambda ,\chi(z)\right) 
    \label{modif_phi}
\end{equation}
where $\langle \delta^2_{\rm true} \rangle$ is the true variance of slices of the density field at redshift $z$ and of radius $\chi(z)\theta$, while $\langle \delta^2_E \rangle$ is to be understood as the variance that results from the reconstruction scheme (as schematically done in equation~\eqref{eq::schematic_sigma} for the whole projection) and whose modelling we describe in the next subsection.

\subsection{Reconstructed variance along the line-of-sight}

We now take a more detailed look at the $\langle \delta^2_E \rangle$ term that appears in equation~\eqref{modif_phi}. In the Limber approximation, it corresponds to a variance measure\footnote{It formally has a unit of length (Mpc or Mpc/h) because the derivation of the projection formula used in equations~\eqref{projection} and \eqref{modif_phi} makes use of cylinders of infinite length $L$ (and not 2D slices) which cancels out in the end result but for which the variance is formally $\langle \delta^2_E \rangle/L$.} of a 2D matter density contrast slice of radius $\chi(z)\theta$ at redshift $z$ which acts locally as the (reconstructed) convergence from a single lens plane. 
In the absence of a mask, it can be written as 
\begin{equation}
    \left\langle\delta^2_{\rm true}\right\rangle=\sum_{l} \frac{2 l+1}{4 \pi \chi(z)^2} W_{l}\left(\chi(z)\theta\right)^2 P_{\mathrm{NL}}(l / \chi(z),z),
    \label{unmasked_variance}
\end{equation}
where $P_{\rm NL}$ is the nonlinear matter power spectrum 
and $W_l$ is the angular top-hat window function in harmonic space given by
\begin{equation}
    W_l\left(\theta_0\right)=\frac{P_{l-1}\left(\cos \left(\theta_0\right)\right)-P_{l+1}\left(\cos \left(\theta_0\right)\right)}{(2 l+1)\left(1-\cos \left(\theta_0\right)\right)}
\end{equation}
where $P_{l}$ are Legendre polynomial of degree $l$.

In the presence of masks, equation~\eqref{unmasked_variance} is modified as follows \citep{gattiDES}
\begin{equation}
    \left\langle\delta^2_E\right\rangle= \frac{1}{f_{\rm sky}} \sum_{l} \frac{2 l+1}{4 \pi\chi(z)^2} W_{l}\left(\chi(z)\theta\right)^2 f_{l}^{-1} \sum_{l^{\prime}} M_{l l^{\prime}}^{E E, E E} P_{N L}\left(l^{\prime} / \chi(z), z\right) f_{l^{\prime}},
    \label{masked_variance}
\end{equation}
where $M_{l l^{\prime}}^{E E, E E}$ is the element of the mode-mixing matrix that accounts for the transfer of power in the power spectrum from the unmasked shear E-modes to the masked shear E-modes. This term is explained in more detail in appendix~\ref{appendix::masking}. Here, $f_{l} = [(l+2)(l-1)]/[l(l+1)]$ and accounts for the passage from convergence to the shear power spectrum in order to apply the mode-mixing formalism to the shear field directly. Finally, the $f_{\rm sky}$ factor is not necessarily to be understood as the true sky fraction observed by the survey. Instead, it actually comes from the parametrisation of the full-sky reconstructed convergence PDF as a sum of the fluctuations within the mask weighted by $f_{\rm sky}$ and a Dirac-delta (for the unobserved sky regions) weighted by $(1-f_{\rm sky})$. However, this parametrisation, though very accurate in practice, is not exact at the boundaries and holes of the mask, and even more so after smoothing the reconstructed convergence field. Having said that, this $f_{\rm sky}$ term is not free either as it can directly -- without any theoretical input -- be estimated from data as the ratio of the reconstructed smoothed full-sky variance to the one within the considered pixels. Note moreover that the reconstructed $\kappa_E$ unmasked variance values measured after masking from the Takahashi simulation maps tend to be equal to the $\kappa_E$ full-sky variance divided by the true fraction of the full-sky observed in the mask up to less than a percent for the DES Y3 mask and for the range of smoothing scales that we tested (up to $\sim$ 30 arcmin), hinting to the fact that our parametrisation is very accurate.

\section{Testing the model}
\label{sec::test}

In this section we implement the theoretical formalism for the Kaiser-Squires reconstructed convergence PDF described in the previous section and compare it to measurements made in the Takahashi simulations. Given that the simulated shear maps also contain higher-order lensing corrections, the comparison presented here allows us to test that the mask modelling is accurate even when neglecting other sources of B-modes than the ones created by the mask. 
We study two test cases and a third in appendix~\ref{sec::SQUAREmask}. In subsection~\ref{sec::nulled} we test the reconstruction scheme in a regime where the large deviation theory is known to be very accurate, that of the Bernardeau-Nishimichi-Taryuya (BNT) transform \citep{Nulling} which allows us to construct lensing observables only sensitive to the dynamics of the quasi-linear regime of the matter density field. It allows us to understand how well we are probing the Kaiser-Squires reconstruction without mixing additional inaccuracies that the theory could present. In subsection~\ref{sec::DESmask}, we test our theoretical $\kappa_E$ PDF for a source distribution mimicking that of the 4$^{\rm th}$ bin from DES Y3 analysis and where we applied the real DES Y3 mask to the full-sky shear field. We present in appendix~\ref{appendix::KS-moments} supplementary information to subsection~\ref{sec::DESmask} but at the level of the cumulants and in appendix~\ref{sec::SQUAREmask} the effect of a reconstruction similar to subsection~\ref{sec::DESmask} but where we replace the DES Y3 mask by a square patch with no holes and of the same area ($\sim$ $70 \times 70$ deg$^2$).

\subsection{Mock data preparation and simulation-specific corrections}

\begin{figure}
    \centering
    \includegraphics[width = 0.8\columnwidth]{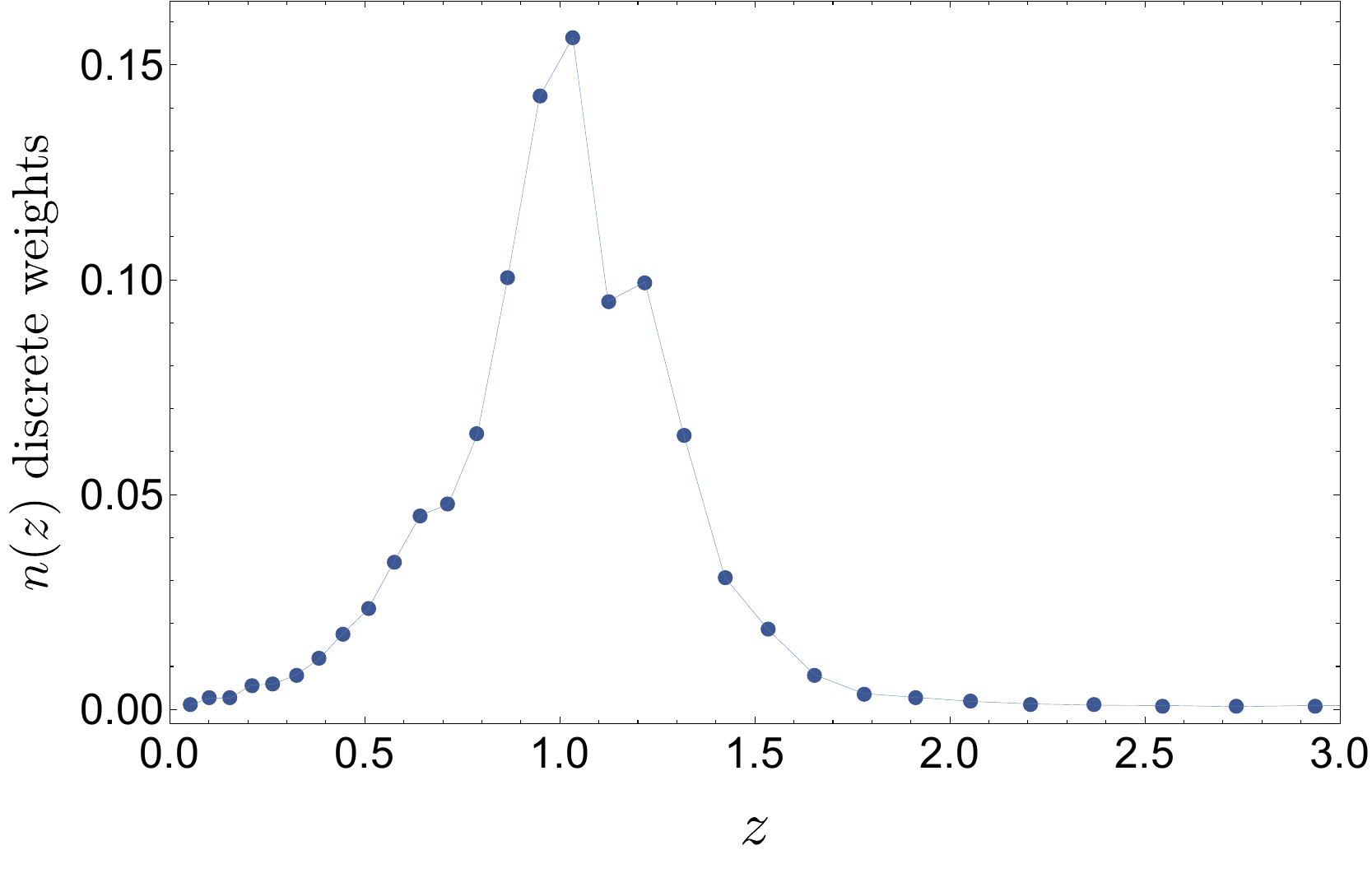}
    \caption{Weights applied to the discrete source planes of the Takahashi simulation to mimic the DES Y3 source distribution in its fourth redshift bin. The weights are normalised so that their sum is equal to one. The thin line is only there to guide the eye and does not have any meaning.}
    \label{n_z}
\end{figure}

As mentioned previously, the simulated true convergence and cosmic shear maps used in this work are obtained from the publicly available N-body simulations by Takahashi et al. \cite{takahashi_full-sky_2017}. The simulation suite provides weak lensing maps for $N=38$ fixed source redshift planes between $z_s = 0.05$ and $z_s = 5.3$ in \verb |Healpix| format. We combine these maps according to a source distribution $n(z)$ inspired by the 4$^{\rm th}$ tomographic bin of the DES Y3 analysis to simulate a DES Y3-like lensing map in its 4$^{\rm th}$ redshift bin. The procedure to do that is as follows:\footnote{Although we show here the procedure on the $\kappa$ field, the same source plane combination scheme is applied to create the shear maps.}
\begin{equation}
    \kappa_{n(z)} = \sum_{i=1}^{N} s_i \kappa_{z_{s}^i} \ ,
\end{equation}
where $s_i$ is a specific weight for a given full-sky lensing map $\kappa_{z_{s}^i}$ at source plane $z_{s}^i$. The weights used for combining the discrete source planes from the simulations are shown in figure~\ref{n_z}. This final simulated convergence map can then be expressed as a line-of-sight projection of the matter density contrast through equation \eqref{def-convergence} with a lensing kernel $w_{n(z)}$:
\begin{equation} 
    w_{n(z)}(z) = \sum_{i=1}^{N} s_i w_{z_{s}^i}(z) \ .
\end{equation}
The $w_{z_{s}^i}$ is in turn the lensing kernel for a given discrete source plane at redshift $z_{s}^i$ and reads
\begin{equation} 
    w_{z_{s}^i}(z) = \frac{3 \Omega_{\mathrm{m}} H_0^2}{2c^2} \frac{[\chi_{z_{s}^i} - \chi(z)]\chi(z)}{\chi_{z_{s}^i}}\mathcal{H}(z_s-z)(1+z) \ .
\end{equation}

\noindent
To fairly compare the theory and our measurement from the simulations, we further include a few corrections to the theoretical modelling described in section~\ref{sec::model}. These corrections are specific to the Takahashi simulation and take into account inaccuracies in the simulation itself rather than additional effects present in a real survey. Firstly, we take into account the fact that the simulation has discrete spherical lens shells of thickness 150 Mpc/h. This is done both by replacing the continuous integrations along the line-of-sight by discrete sums
\begin{equation}
    \int {\rm d}\chi f(\chi) \rightarrow \sum_{i} 150 \times f[150(i-0.5)],
\end{equation}
and at the level of the nonlinear power spectrum by correcting it following (equation (28) in appendix B of ref.~\cite{takahashi_full-sky_2017}):
\begin{equation}
    P_{\rm NL}(k,z) \rightarrow \frac{\left(1+c_1 k^{-\alpha_1}\right)^{\alpha_1}}{\left(1+c_2 k^{-\alpha_2}\right)^{\alpha_3}} P_{\rm NL}(k,z),
    \label{corr1}
\end{equation}
with $c_1 = 9.5171 \times 10^{-4}$, $c_2 = 5.1543 \times 10^{-3}$, $\alpha_1 = 1.3063$, $\alpha_2 = 1.1475$ and $\alpha_3 = 0.62793$. Note that the wave-modes $k$ are in units of h/Mpc and the power spectrum in units of (Mpc/h)$^3$. Finally, in a manner analogous to taking into account the pixel window function at the map level, the resolution effects of the Takahashi maps can be taken into account by multiplying the nonlinear power-spectrum by a damping factor that depends on the {\sc nside} of the \verb |Healpix| map (equation (5) in ref.~\citep{takahashi_full-sky_2017}):
\begin{equation}
    P_{\rm NL}(l/\chi(z),z) \rightarrow \frac{P_{\rm NL}(l/\chi(z),z)}{1+(l/1.6/\text{\sc nside})^2} \ .
    \label{corr2}
\end{equation}
Note that taking into account corrections~\eqref{corr1} and \eqref{corr2} at the level of the nonlinear power spectrum has some effects on all other higher-order (density contrast) cumulants through their scaling relations with the variance while keeps the $S_n$ ratios constant in equation~\eqref{Sp}.

\subsection{DES Y3 mask and BNT transform}
\label{sec::nulled}

One issue faced by theoretical approaches that aim at describing quantities projected along the line-of-sight is the mixing of both very nonlinear scales not accurately probed by standard from-first-principles 
approaches such as ours, and reasonably larger (quasi-linear) scales more accessible to the theory. As such, usual weak lensing statistical probes are (i) modelled accurately only on sufficiently large angular scales with scale-cuts on small angular scales, so as to mitigate the influence of small nonlinear physical scales at the tip of the light-cone, (ii) modelled by more phenomenological approaches such as halo models that can also take into account nonlinear and baryonic physics which becomes important on small scales \citep{Mead_2021_HMCode}, (iii) or by making use of numerical recipes and simulations for e.g. specifically incorporating baryonic feedback effects \citep{Baryonification} which have unfortunately not been tested in great detail, especially for higher-order non-Gaussian statistics. 

Alternatively, a theoretical strategy to disentangle quasi- and non-linear scales in lensing quantities known as the Bernardeau-Nishimichi-Taruya (BNT) transform or nulling strategy was proposed by \cite{Nulling}. It allows for very accurate theoretical predictions in the context of power spectrum analysis and has recently been extended to the convergence and aperture-mass PDFs \citep{Barthelemy20a,Barthelemy21}. This nulling strategy was used recently in \cite{x-cut} to remove the sensitivity to the poorly modelled small scales for the cosmic shear 2PCF, and therefore improve cosmological constraints using the DES Y1 shear data. This strategy could become even more relevant for future lensing experiments with better knowledge of redshifts and the division of sources in more redshift bins.

This BNT transform can only be used in the context of a tomographic analysis of at least 3 source redshifts (or redshift bins, although not treated here) and is a linear transformation $M$ applied to the set of lensing kernels $\omega_i \equiv \omega(\chi,\chi_{s,i})$, giving rise to a new set of re-weighted kernels 
\begin{equation}
    \Tilde{\omega}^j = M^{ij}\omega_i.
\end{equation}
For a set of 3 source planes labeled from $j= i-2$ to $j = i$ arranged by ascending order, it was shown in \cite{Nulling} that $M$ must satisfy the system
\begin{equation}
\left\{ \begin{aligned}
        &\sum_{j=i-2}^{i} M^{j i}=0, \\
        &\sum_{j=i-2}^{i} \frac{M^{j i}}{\chi_{s,j}}=0 \ .
        \end{aligned}
\right.
\end{equation}
This system of equations is under-constrained and hence we also impose by convention $M^{ii} = 1$. The elements of $M$ can thus be computed considering sequential triplets of tomographic bins, going from the lowest to the highest redshift, such that
\begin{align}
    M^{i-2,i} = \frac{\chi_{i-2}(\chi_{i-1}-\chi_{i})}{\chi_{i}(\chi_{i-2}-\chi_{i-1})},\\
    M^{i-1,i} = \frac{\chi_{i-1}(\chi_{i}-\chi_{i-2})}{\chi_{i}(\chi_{i-2}-\chi_{i-1})}.
\end{align}

\begin{figure}
    \centering
    \includegraphics[width = 0.8\columnwidth]{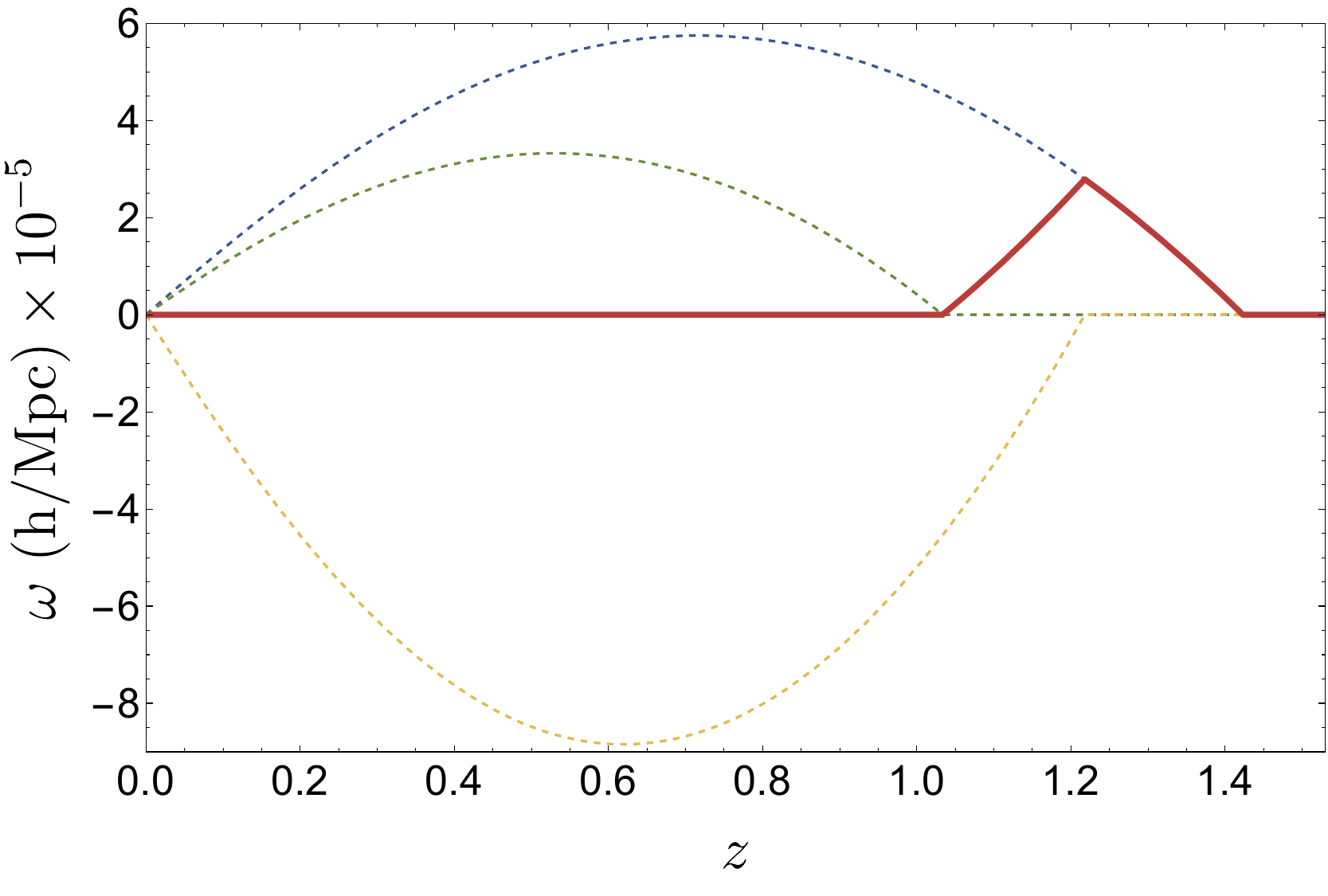}
    \caption{Illustration of the effect of the BNT transform on lensing kernels. The green, yellow and blue dashed lines are the kernels up to $z_s = 1.0, 1.2, 1.4$ respectively re-weighted by their appropriate BNT coefficients, $M^{ij} = [0.80, -1.80, 1]$ where $j$ is fixed and equal to 3 if the blue kernel is the third of a tomographic analysis. The thick red line is the sum of the 3 re-weighted kernel. The effect of nulling is to set to zero the contribution of all lenses below the closest source plane.}
    \label{null_kernel}
\end{figure}

We display in figure~\ref{null_kernel} an example for a set of 3 source planes located at $z_s = 1.0, 1.2, 1.4$. This is the kernel we use in this subsection. The green, yellow and blue dashed lines are the kernels up to $z_s = 1.0, 1.2, 1.4$ respectively re-weighted by their appropriate BNT coefficients while the thick red line is the sum of the 3 re-weighted kernels. Note that the blue dashed line is also the original kernel since its BNT coefficient is set to 1. One can thus clearly see that the effect of nulling is to set to zero the contribution of all lenses below the closest source plane and therefore remove the contribution of small scales (at the tip of our light-cone) which are very nonlinear and where the effect of baryonic physics becomes non-negligible.

For our purpose, the BNT transform -- which boils down to a simple linear combination of the maps -- is applied at the level of the masked shear field\footnote{Since the BNT transform is a linear transformation, it can without distinction be applied at the level of the (masked) shear or the convergence. The consideration of nonlinear high-order lensing or reduced shear corrections would formally break the exactness of the nulling but would still mitigate the influence of small scales so that it could still be used on real data. The same comment applies to the inexact knowledge of the underlying cosmology: an inexact nulling still mitigates the influence of small scales.} and is also straightforward to implement in our theoretical approach to the convergence PDF since we only need to replace the original kernel with its nulled counterpart. 

\begin{figure}
    \hspace{-1.1cm}
    \begin{subfigure}{0.48\columnwidth}
        \hspace{-0.9cm}
        \includegraphics[width=1.25\columnwidth]{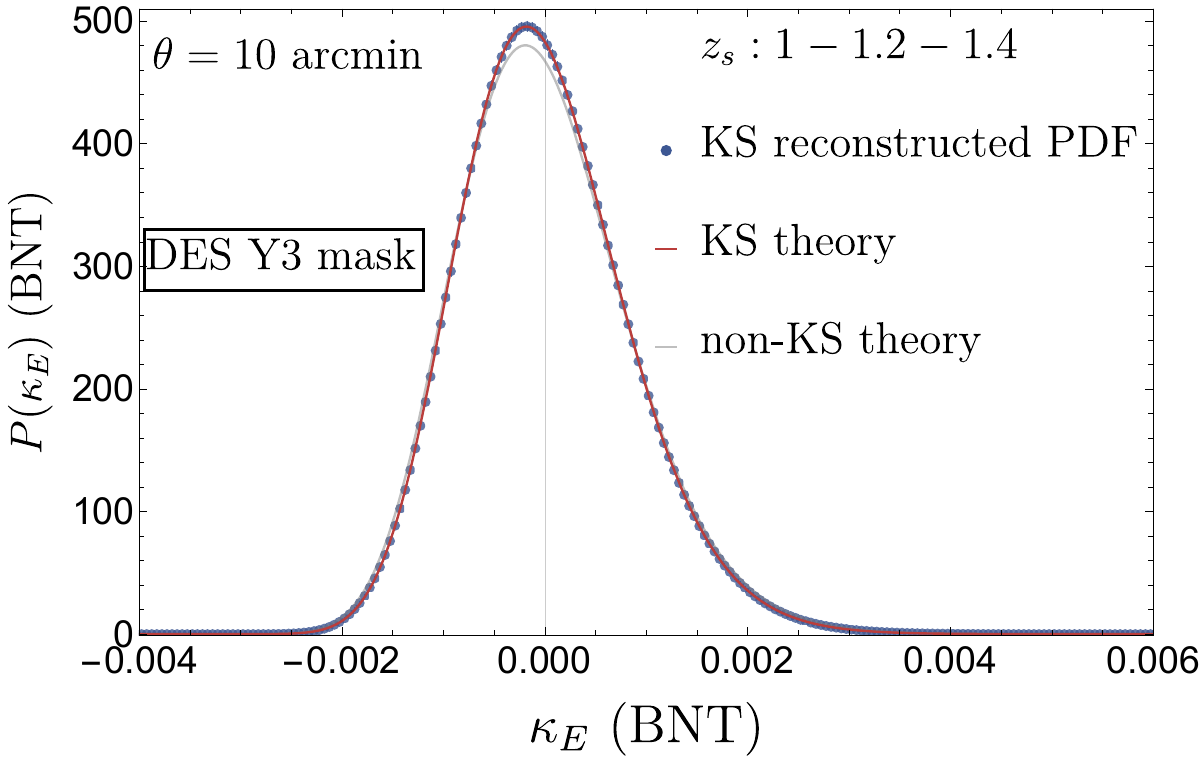}
    \end{subfigure}
    \hfill
    \begin{subfigure}{0.48\columnwidth}
        \hspace{-0.3cm}
        \includegraphics[width=1.25\columnwidth]{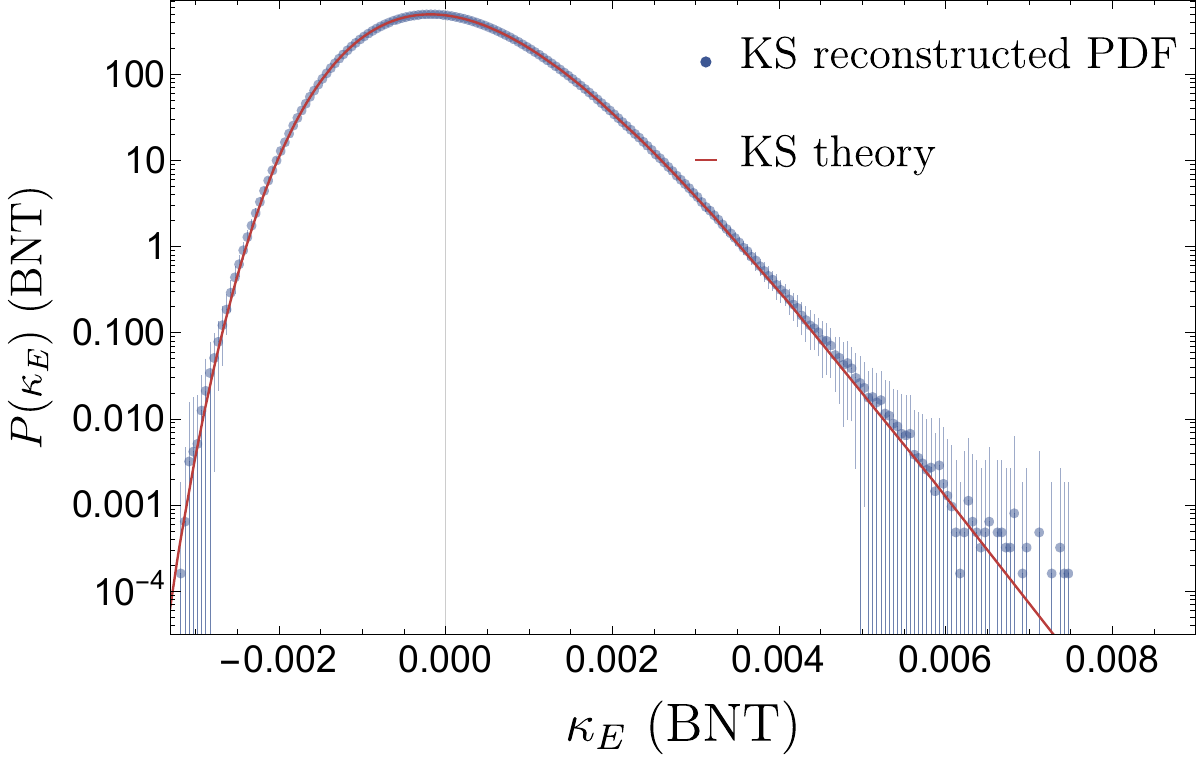} 
    \end{subfigure}
    \begin{subfigure}{\columnwidth}
    \centering
        \includegraphics[width=1.05\columnwidth]{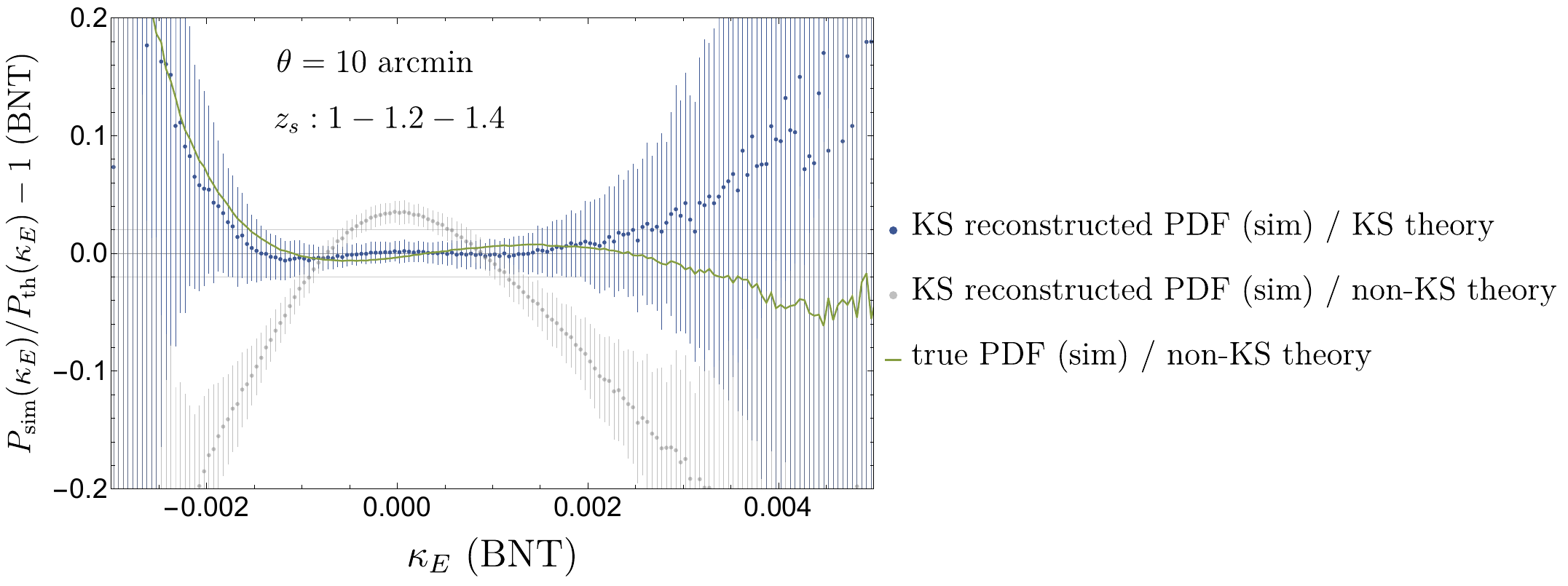} 
    \end{subfigure}
    \caption{\textit{Top panels:} PDF of the Kaiser-Squires reconstructed nulled convergence map using source planes located at $z_s =$ 1.0, 1.2 and 1.4 under the DES Y3 mask. The red line is the theory described by equation~\eqref{modif_phi} and the blue points with error bars are the mean and $1\sigma$ fluctuations of the measured PDF in 108 realisations. The grey solid line is the theory described by equation~\eqref{projection} that is without taking into account the KS reconstruction masking effect. \textit{Bottom panel:} Residuals between the theory and the simulation. The green line is the residual of the original non-KS theory to the true, non-reconstructed simulated convergence PDF. It is highly compatible with the residuals between the KS-theory and the simulated reconstructed KS PDF (blue points). Neglecting the KS reconstruction effects (grey points) significantly worsens the quality of the theoretical prediction while taking them into account does not improve nor worsen the predictive power of our LDT formalism.}
    \label{DES_null}
\end{figure}

We then show in figure~\ref{DES_null} -- in the regime where the traditional BNT transformed PDF is perfectly described by the LDT formalism -- how the theoretical integration of the convergence reconstruction scheme performs. The BNT convergence field is smoothed with a top-hat window of radius $\theta = 10$ arcmin. By looking at the top panels, one can appreciate that the exponential cut-off in the tails of the PDF, a prediction of our formalism, is well-observed once one reduces the lensing kernel down to scales accessible to from-first-principles theoretical modelling (such as ours). More interestingly, this behaviour is still observed through the Kaiser-Squires reconstruction scheme presented in section~\ref{sec::KS}, and our theoretical modelling of this reconstruction does not reduce our ability to capture the shape of the PDF in this regime. This hints at the fact that our model is accurate enough to reproduce the reconstruction effects, at least in the regime where our initial large-deviation formalism is accurate. This is further illustrated in the bottom panel of figure~\ref{DES_null} where blue points denote the residuals of our KS theory with respect to the measurement from the KS reconstructed simulations. The green line describes the residual between the original LDT theory without reconstruction and the measurement of the PDF from the true $\kappa$ map (i.e.~not reconstructed from the shear field using KS method). Here we observe that the two residuals are highly compatible. Moreover, we also find that taking the reconstruction into account does not reduce in the slightest our ability to describe the PDF in this regime, while neglecting it (grey points) would seriously damage our predictive power.

\subsection{DES Y3 mask for sources in the fourth DES Y3 redshift bin}
\label{sec::DESmask}

\begin{figure}
    \hspace{-1.1cm}
    \begin{subfigure}{0.49\columnwidth}
        \hspace{-0.9cm}
        \includegraphics[width=1.25\columnwidth]{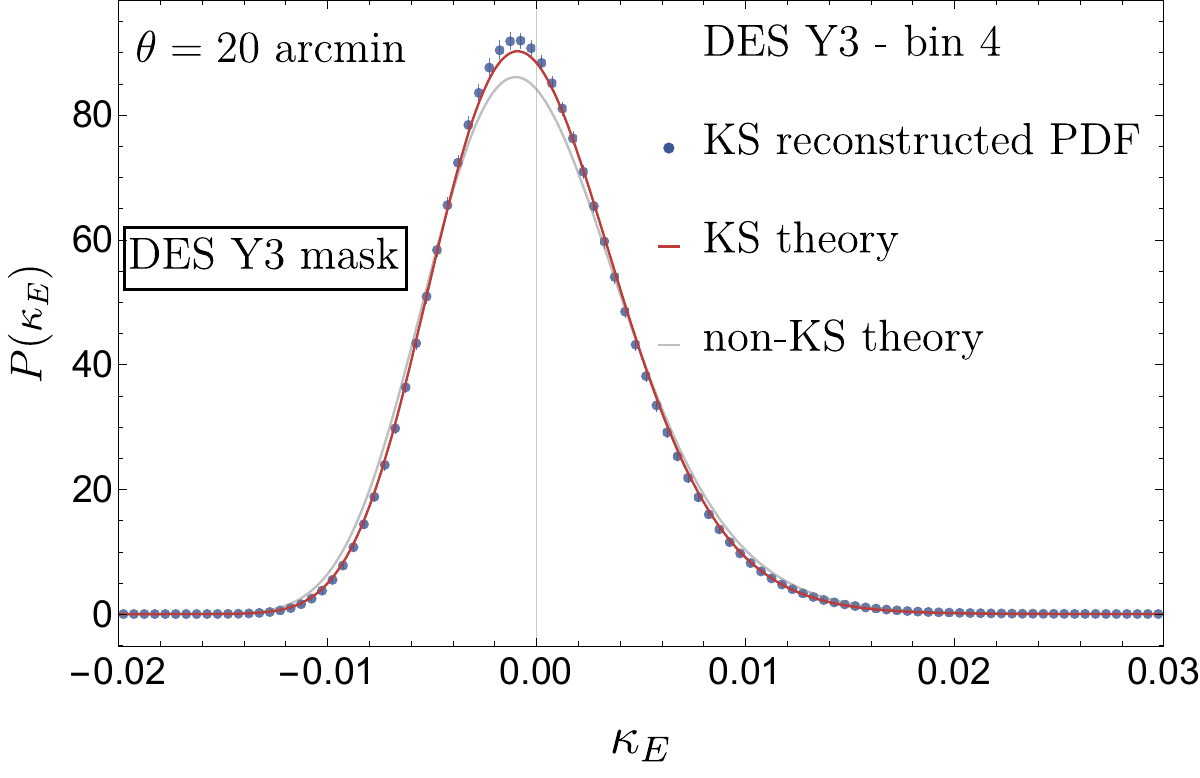}
    \end{subfigure}
    \hfill
    \begin{subfigure}{0.49\columnwidth}
        \hspace{-0.3cm}
        \includegraphics[width=1.28\columnwidth]{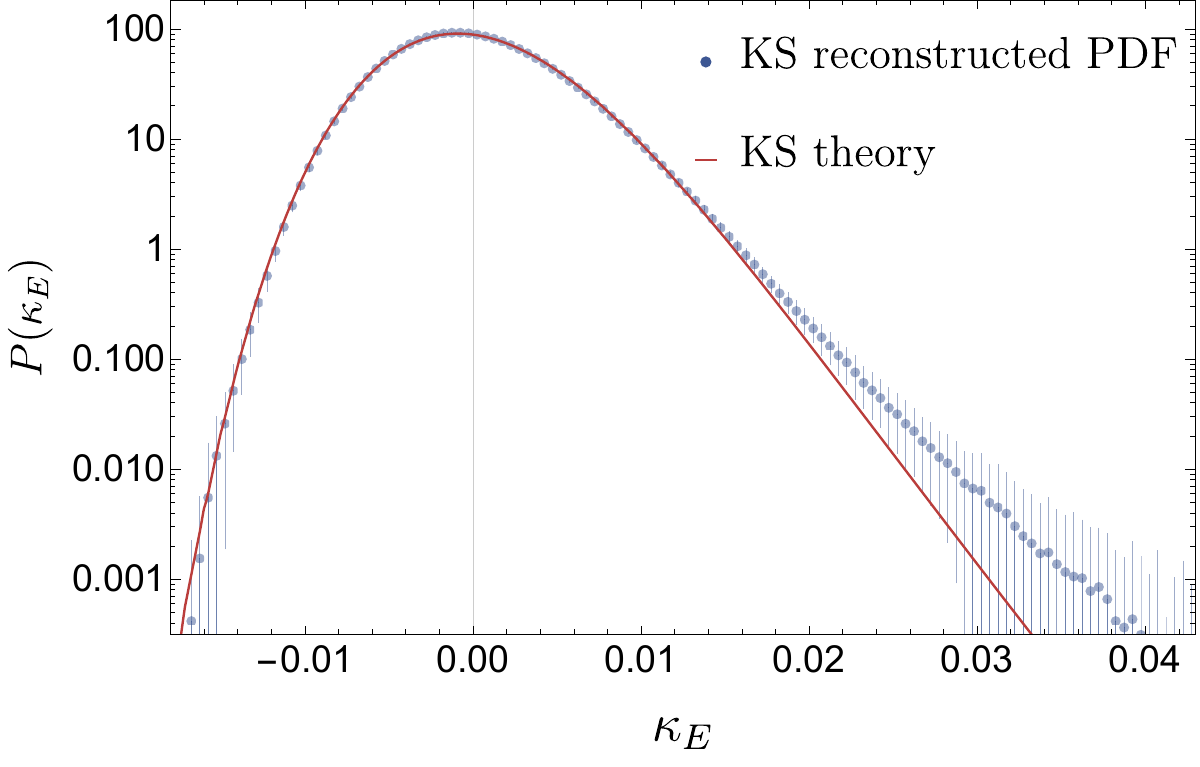} 
    \end{subfigure}
    \begin{subfigure}{\columnwidth}
    \centering
        \includegraphics[width=1.05\columnwidth]{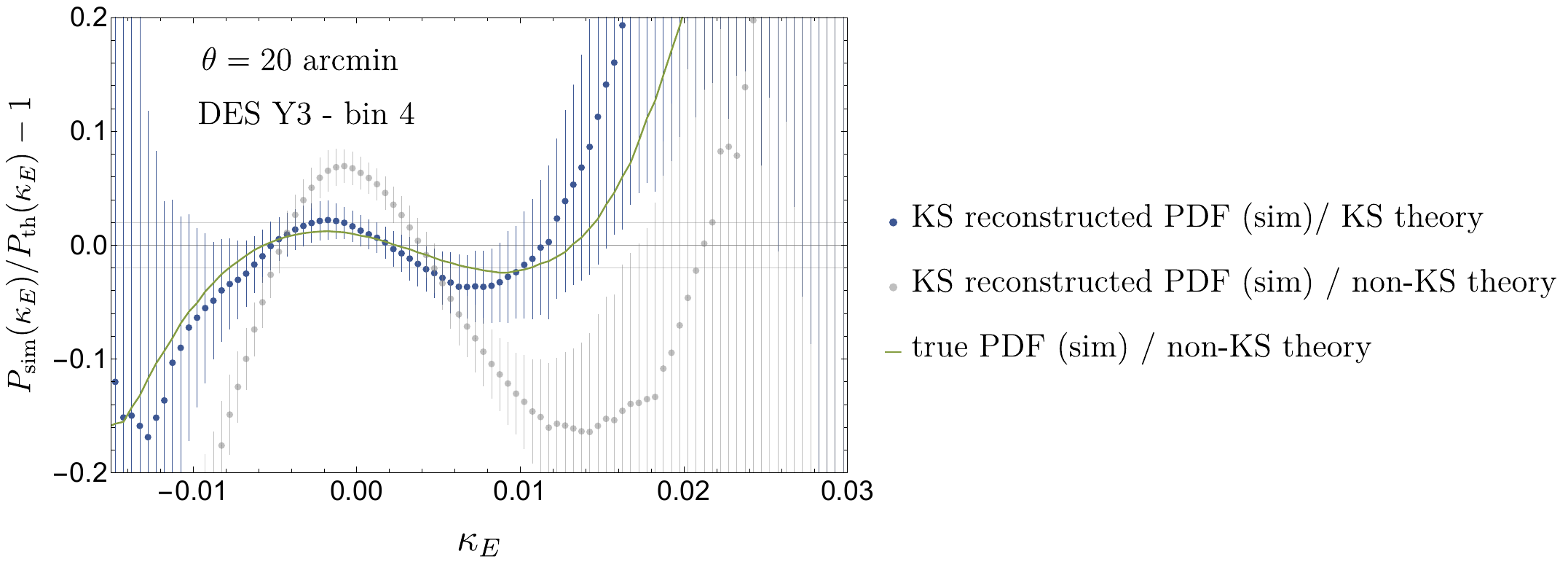} 
    \end{subfigure}
    \caption{\textit{Top panels:} PDF of the Kaiser-Squires reconstructed convergence map for the simulated DES Y3-like 4$^{\rm th}$ bin in the Takahashi simulation and under the DES Y3 mask. The red line is the theory described by equation~\eqref{modif_phi} and the blue points with error bars are the mean and $1\sigma$ fluctuations of the measured PDF in 108 masked Takahashi maps. The grey solid line is the theory described by equation~\eqref{projection} that is without taking into account the KS reconstruction. \textit{Bottom panel:} Residuals between the theory and the simulation. The green line is the residual of the original non-KS theory to the true, non-reconstructed simulated convergence PDF. It is highly compatible with the residuals between the KS-theory and the simulated reconstructed KS PDF (blue points). Neglecting the KS reconstruction effects (grey points) significantly worsens the quality of the theoretical prediction while taking them into account preserves the predictive power of our LDT formalism. We note that the slight increase in the size of residuals is due to the complicated shape of the mask and presence of holes as demonstrated in figure~\ref{square_zs1} for the case of the square mask of similar area.}
    \label{DES_zs1}
\end{figure}

For the more traditional case of a source distribution mimicking the 4$^{\rm th}$ source bin of the DES Y3 analysis (that we replicate using the source redshift distribution shown in figure~\ref{n_z}), and for a shear field under the DES Y3 mask, we show in figure~\ref{DES_zs1} the results of our theoretical prediction for the reconstructed $\kappa_E$ PDF. The field is smoothed with a top-hat window of radius $\theta = 20$ arcmin. By looking at the top-left panel, one can appreciate that the theoretical formalism seems to capture well the bulk of the PDF when considering the $1\sigma$ fluctuations across the 108 Takahashi realisations estimating the diagonal of the PDF covariance matrix. This is to be contrasted with the theoretical formalism that neglects the KS reconstruction effects and thus performs poorly which justifies the need for its modelling. Looking at the PDF on a log-scale (top-right panel), one can however observe an increasingly larger departure of the theory from the simulation as one enters the high and low-density tails. This is expected since cumulants from the large-deviations formalism are only valid in the quasi-linear regime and the mixing of scales (that the BNT transform would prevent as shown in subsection~\ref{sec::nulled}) worsens the quality of the theoretical prediction in the tails.

This is clearly seen in the residuals between the theory and the simulation (bottom panel) where a typical mismatched skewness modulation\footnote{Let us remind here that the skewness enters the Edgeworth expansion of the PDF at the first non-Gaussian correction order and multiplies a third order Hermite polynomial of the convergence field as follows
\begin{equation*}
{\mathcal P}(\kappa)={\cal G}(\kappa)\left[1+\sigma \frac{S_{3,\kappa}}{3!} H_3\left(\frac{\kappa}{\sigma}\right) +{\cal O}(\sigma^2)\right],
\end{equation*}
where $H_3(x)=x^3-x$. We emphasise that the predictions shown in this paper do not use or assume an Edgeworth expansion. It is however useful to interpret the shape of the residuals between the theory and the simulation.} is visible, showing that higher order correction to the skewness would be needed if we had smaller error bars, especially since only the cosmic variance is here used to obtain those error bars. The alternatives include looking at larger angular scales, considering sources at higher redshifts, incorporating modelling errors in the error budget or making use of the BNT transform shown in subsection~\ref{sec::nulled} which we consider the most principled approach.

Focusing on the accuracy of the modelling of the KS reconstruction, we now look at the green line in the bottom panel which describes the residual of the original LDT theory without KS reconstruction to the measurement of the PDF from the true $\kappa$ map (i.e.~not reconstructed from the shear field using KS method). As in the previous subsection, we once again observe that the two residuals are highly compatible, highlighting the fact that the reconstruction does not weaken nor improve our ability to describe the PDF in this regime. This again hints at the fact that our implementation of the KS effects on the PDF is accurate enough.

\section{Incorporating observational systematics in the modelling}
\label{sec::systematics}

In this section we discuss the modelling of additional effects in the reconstructed $\kappa$-PDF, including galaxy intrinsic alignments, baryonic feedback effects, additive and multiplicative shear biases, higher-order lensing corrections and photometric redshift uncertainties.

\subsection{Intrinsic alignments}
\label{section::IA}

As noted in equation~\eqref{eq:observed_ellipticity},
the observed source galaxy ellipticity is a combination of the gravitational lensing shear component $\gamma$, the intrinsic ellipticity of the galaxies $\epsilon_{\rm IA}$ induced by correlations with local gravitational tidal fields at the source, and the random stochastic component that contributes as shape noise. In this section we include the $\epsilon_{\rm IA}$ (also known as intrinsic alignment) component in our framework using the popular \textit{non-linear tidal alignment} (NLA) model \citep{2007MNRAS.381.1197H, 2007NJPh....9..444B}. Indeed, recent work based on a perturbative field-level forward modelling of weak lensing fields \citep{Porqueres2023} indicates that on scales of about $15$ arcmin, the NLA model is adequate and leads to conservative and unbiased cosmology constraints even when analysing data generated through a more complex tidal alignment and tidal torquing (TATT) intrinsic alignment model \citep{Blazek2019_TATT}.

As shown in section~\ref{sec::gamma2kappa}, through second derivatives of the projected gravitational potential $\psi$, the lensing shear $\gamma$ is related to the lensing convergence $\kappa$, the latter being expressed as a line-of-sight integration of the matter density contrast $\delta$ (see equation~\eqref{def-convergence}). In analogy, for the intrinsic ellipticity component $\epsilon_{\rm IA}$ we can define a quantity $\kappa_{\rm IA}$
\begin{equation}
    \kappa_{\rm IA}({\bm \vartheta}) \equiv \int_0^{+\infty} {\rm d}z \frac{{\rm d}\chi}{{\rm d}z} \, n(z) \, \delta_{\rm IA}(\chi(z),\chi(z){\bm \vartheta}),
    \label{def-convergence_IA}     
\end{equation}
where $\delta_{\rm IA}$ is a three-dimensional field that \textit{effectively} determines the intrinsic alignment (IA) of the source galaxies with their local gravitational tidal fields. Note that the line-of-sight projection kernel in equation~\eqref{def-convergence_IA} is the source galaxy redshift distribution $n(z)$, and not the lensing kernel $\omega_{n(z)}(z)$ as in equation~\eqref{def-convergence}.

In the NLA model, $\delta_{\rm IA}$ can be expressed as a first order bias expansion with the nonlinear matter density contrast field\footnote{Strictly speaking, in the \textit{linear tidal alignment} model one performs the bias expansion around the linear matter density contrast field $\delta_{\rm lin}$ (as is correct in perturbation theory), whereas in the NLA model one simply replaces $\delta_{\rm lin}$ with the nonlinear matter density field $\delta$ (as adopted in our work).} \citep{2007MNRAS.381.1197H,2007NJPh....9..444B}:
\begin{equation}
    \delta_{\rm IA}(\chi,\chi{\bm \vartheta}) = f_{\rm IA}(\chi(z)) \ \delta (\chi,\chi{\bm \vartheta}) \ .
    \label{def-NLA}
\end{equation}
The $f_{\rm IA}(z)$ term is the amplitude of the intrinsic alignment of the specific source galaxies and is given by
\begin{equation}
    f_{\rm IA}(z) = -A_{\text{IA}}\left(\frac{1+z}{1+z_0}\right)^{\alpha_{\rm IA}}\frac{c_1\rho_{\text{crit}}\Omega_{\text{m}}}{D(z)} \ ,
    \label{def-f_IA}
\end{equation}
where $A_{{\rm IA}}$, $\alpha_{\rm IA}$ are redshift-independent parameters, $c_1 = 5\times10^{-14}$  $(h^2 M_{\odot}/{\rm Mpc^3})^{-1}$\citep{2002MNRAS.333..501B}, $\rho_{\rm crit}$ is the critical energy density, $D(z)$ is the growth factor normalised to unity today, and $z_0$ is a pivot redshift. The fact that $c_1$ was calibrated in $h^2$ units allows us to fix $c_1\rho_{\text{crit}} = 0.0134$. Therefore, the intrinsic alignment of source galaxies can thus be readily included in the overall convergence signal as an additive component on top of the lensing $\kappa$
\begin{equation}
    \kappa \rightarrow \kappa + \kappa_{\rm IA} \ .
\end{equation}
In practice, the inclusion of the intrinsic alignment effect in the modelling of the PDF $P(\kappa)$ within the NLA framework is straightforward and can be achieved by replacing the lensing kernel in equation~\eqref{def-convergence} with 
\begin{equation}
    \omega_{n(z)}(z)\rightarrow \omega_{n(z)}(z)+n(z)f_{\rm IA}(z) \ .
    \label{eq:correction_IA}
\end{equation}

One can treat these intrinsic alignment terms as nuisance parameters which can be marginalised over in order to constrain cosmological parameters of interest when analysing $\mathcal{P}(\kappa)$. Note that even though marginalized constraints over the IA parameters in the DES Y3 shear 2PCF analysis are compatible with no IA (see for example \cite{2022PhRvD.105b3514A} or \cite{2022PhRvD.105b3515S}), neglecting the IA effect would bias the constraints on other parameters of interest. Based on those analyses, the typical order of magnitude of the IA effect would be around $A_{IA}\sim 0.4$ and $\alpha_{IA} \sim 1.7$ but again one should not fix those values in an analysis, also because the uncertainty on the IA nuisance parameters is one of the limiting aspect of cosmic shear analysis that one might want to mitigate through the use of high-order statistics. For the PDF, the interplay with cosmological parameters will be studied in more detail in our upcoming work (Paper II in the series). One potentially important follow-up line of work would consist in extending the current bias-like IA expansions to the PDF so that higher-order terms in models such like TATT to alleviate for potential biases in the IA treatment. Current DES Y3 analysis of convergence moments \cite{gattiDES} or third-order aperture mass in KiDS-1000 \cite{kids-map3} which both contain NLA modelling are nevertheless respectively compatible with the main 3x2 points analysis from DES Y3 which is made with TATT \cite{2022PhRvD.105b3515S} and the KiDS-1000 2-point analysis \cite{kids-map3}.

\subsection{Baryonic feedback}

\begin{figure}
    \centering
    \includegraphics[width=0.8\columnwidth]{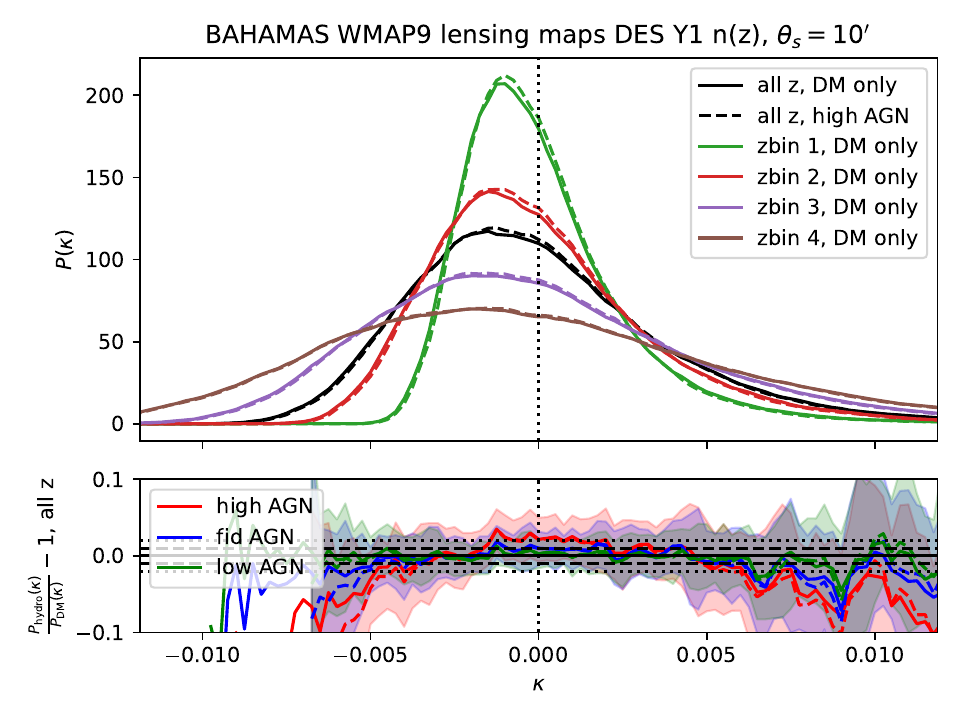}

    \caption{\textit{Top panel:} Smoothed convergence PDF from the BAHAMAS runs for the full DES Y1 $n(z)$ (black) and the 4 tomographic bins (colours) for the DM-only run (solid) and the high AGN run (dashed) averaged over 25 lensing cones. The dotted vertical line at $\kappa=0$ highlights the non-Gaussian shape of the $\kappa$-PDF.
    \textit{Bottom Panel:} Residual of convergence PDF comparing the hydro BAHAMAS runs with fiducial (blue), high (red) and low (green) AGN feedback to the DM-only result for the whole $n(z)$. The solid lines indicate the residual between the means (averaged over 25 runs) while the shaded bands indicate the mean and standard deviation of the residual ratio in individual runs (obtained from 25 runs). The horizontal black dashed and dotted lines indicate a residual of 2\% and 1\%, respectively.}
    \label{fig:kappaPDF_BAHAMAS}
\end{figure}

\begin{figure}
\centering
\includegraphics[width=0.8\columnwidth]{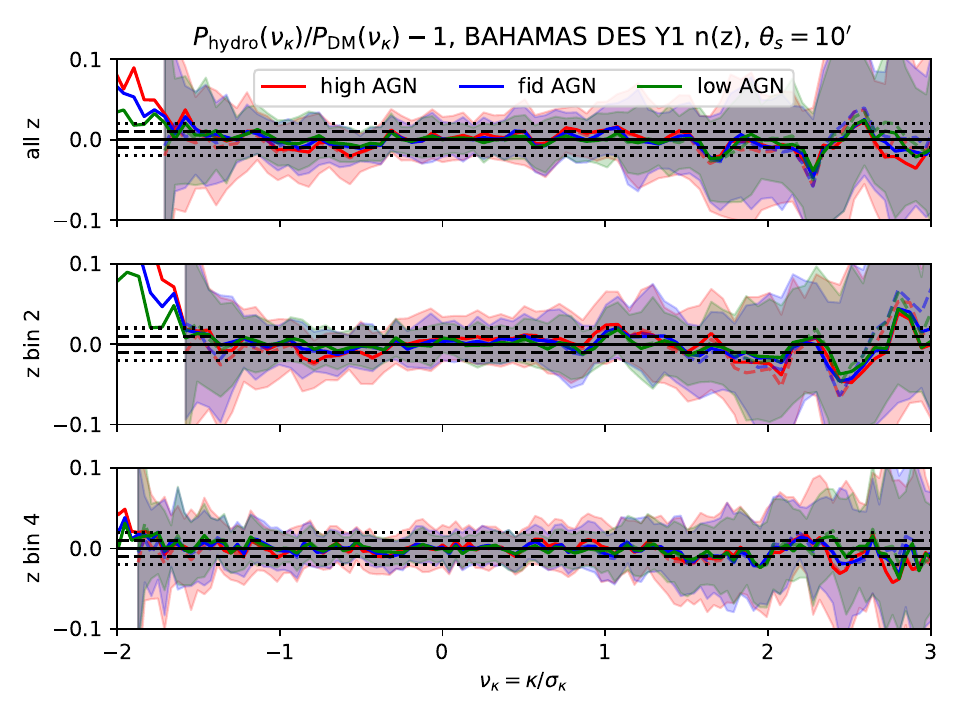}
\caption{Residual of $\nu=\kappa/\sigma$ lensing PDF comparing the hydro BAHAMAS runs with fiducial (blue), high (red) or low (green) AGN feedback to the DM-only result averaged over 25 lensing cones
smoothed for the whole $n(z)$ and in two different tomographic bins corresponding to DES Y1 $n(z)$. Bin 2 corresponds to the range $z\in[0.43,0.63]$ while bin 4 corresponds to $z\in[0.9,1.3]$. The solid lines indicate the residual between the means (averaged over 25 runs) while the shaded bands indicate the mean and standard deviation of the residual ratio in individual runs.}
\label{fig:kappavssigPDF_BAHAMAS_res}
\end{figure}

We determine the effect of baryonic feedback on the mildly nonlinear convergence PDF by relying on simulated DES Y1-like lensing maps from the BAHAMAS hydrodynamical simulation suite \citep{McCarthy_2017_BAHAMAS}, where the strength of AGN feedback has been varied in various simulation runs. The convergence PDF for the full DES Y1 $n(z)$ and the separate redshift bins is illustrated in the upper panel of figure~\ref{fig:kappaPDF_BAHAMAS}. Note that for our purpose, the distribution of sources between DES Y1 and Y3 is similar enough, specifically, the four bins roughly peak respectively at redshift $z_s =$ 0.3, 0.5, 0.8 and 1. In the lower panel we show the residual between the PDFs including baryonic feedback and the dark-matter only runs, which shows a clear signature of a changed standard deviation, whose values we summarise in Table~\ref{tab:sig_BAHAMAS}. Conjecturing that the main impact is on the variance, we obtain the PDFs of the weak lensing convergence divided by its variance $\nu_\kappa=\kappa/\sigma_\kappa$ and show exemplary results for the whole $n(z)$ and two redshift bins in figure~\ref{fig:kappavssigPDF_BAHAMAS_res}. The excellent agreement of the PDFs for the standard deviation-normalised convergence field demonstrates that modelling the impact of baryons at the level of the variance is likely sufficient, as assumed in the recent analysis of HSC Y1 data \cite{2023arXiv230405928T}. Using conservation of probability, this implies that the $\kappa$-PDF in the presence of baryonic feedback is given by
\begin{equation}
    \mathcal P_{b}(\kappa_b) 
    = \mathcal P_{\rm DM}\left(\kappa=\kappa_b\frac{\sigma_{\rm DM}}{\sigma_b}\right)\frac{\sigma_{\rm DM}}{\sigma_b} \,.
    \label{eq:PDF_baryons}
\end{equation}

For the purpose of obtaining the variance correction factor, the baryonic feedback model within HMcode can be used as illustrated in figure~5 of \cite{Mead_2021_HMCode} showing a simple single-parameter model reproducing the 3D matter power spectrum in BAHAMAS.

\begin{table}
    \centering
    \begin{tabular}{|l|c|c|c|c|}
    \hline
        $\sigma_\kappa\ [10^{-3}]$ & DM only & fid AGN & high AGN & low AGN \\\hline
        full $n(z)$ & 3.955 & 3.919 [-0.9\%] & 3.874 [-2.0\%]&  3.944 [-0.27\%] \\
         z-bin 2 [0.43,0.63] & 3.483  & 3.442 [-1.2\%] & 3.398 [-2.4\%] & 3.468 [-0.45\%]\\
         z-bin 4 [0.9,1.3] & 6.149 & 6.129 [-0.33\%] & 6.080 [-1.2\%] & 6.154 [+0.08\%]\\
    \hline
    \end{tabular}
    \caption{Impact of bayronic effects on the standard deviation of the weak lensing convergence at scale $\theta=10'$ for a DES Y1  $n(z)$ as measured from the BAHAMAS simulations}
    \label{tab:sig_BAHAMAS}
\end{table}

We have further checked (not shown) that our the results are fully compatible with similar measurements from kappaTNG maps for single source redshifts \citep{Osato2021kappaTNG} obtained from the IllustrisTNG hydrodynamical simulations \citep{Springel_2018_IllustrisTNG}, and for tomographic bins following the Euclid $n(z)$ as adopted by the Magneticum simulations \citep{Castro2017,Harnois_Deraps_2021}. 

\subsection{Additive and multiplicative shear biases}

We adopt the modelling of any biases coming from the shear measurement pipeline, such as noise bias (e.g. \cite{2012MNRAS.427.2711K}), model-fitting bias (e.g. \cite{2010MNRAS.404..458V}), selection bias (e.g. \cite{2003MNRAS.343..459H}) and bias from the imperfect correction of the image point spread function (PSF; e.g. \cite{2008A&A...484...67P}), with a multiplicative factor $1+m$ to any instance of the estimated shear as is common in literature. As the weak lensing shear and convergence fields are connected to each other via linear transformations, we can propagate the shear measurement biases to convergence with the following transformation using a multiplicative $1+m$ and an additive bias $c$ term
\begin{equation}
    \kappa \longrightarrow (1+m) \kappa + c \ .
\end{equation}
This linear relation holds in the weak lensing regime where $\kappa$ is small. It is common in the literature \citep{2022MNRAS.509.3371M, gattiDES, 2022PhRvD.105b3515S} to consider that these biases are redshift and scale independent within a given source redshift distribution $n(z)$, and thus fixed at the map level. In that case, since the KS reconstruction of the convergence forces us to fix the $l = 0$ wave mode to zero as a straightforward consequence of the mass-sheet degeneracy, we are forcing the mean $\kappa_{(E/B)}$ to be zero which renders inconsistent the consideration of an additive bias $c$. This is consistent with the assumption that any additive bias component can be perfectly removed from the measurement pipeline for compactness through calibration with image simulations on average \citep{2019MNRAS.482..402G}. The PDF of the multiplicative biased shear $\tilde \kappa=(1+m)\kappa$ is straightforwardly obtained by conservation of probability as
\begin{equation}
\mathcal P(\tilde \kappa)=\mathcal P\left(\kappa=\tilde\kappa/(1+m)\right)/(1+m)\,.
    \label{eq:PDF_multbias}
\end{equation}
Previous works have indicated that the presence of shear biases enhances the complementarity of the shear 2PCF and the convergence PDF \citep{Patton17} which may help to further lift parameter degeneracies appearing at the 2-point level.

\subsection{Higher-order lensing corrections}
\label{sec::higher-order-corrections}
At the scales and redshift where the large-deviations formalism can be considered accurate enough to be applied to real data analysis, the corresponding weak-lensing PDFs of either the convergence or the aperture mass can be generated from a finite set of cumulants
in the sense that a correct variance, skewness, and a consistent manner to generate higher-order cumulants yield good results \citep{Barthelemy20a,Barthelemy20b,Barthelemy21}. In that context, studying the corrections to the PDF induced by higher-order lensing corrections amounts to computing the leading-order corrections on the variance and skewness which can then be incorporated straightforwardly to the formalism for the non-linear variance, and by slight modification of the spherical collapse $\nu$ parameter in equation~\eqref{collapse} to match the new values of the skewness along the line-of-sight \citep[see for example][that does it for post-Born corrections in the LDT context]{Barthelemy20b}. We detail in the following the corrections that could be considered in our theoretical formalism and explain why we could discard them for our DES Y3 analysis of the reconstructed $\kappa_E$ PDF. This is consistent with what was done for the weak-lensing moments analysis in DES \citep{GattiDESsim,gattiDES}.

Among all possible corrections, most of the traditional ones have been estimated in past works and in the LDT context. More precisely, the relaxation of the Born approximation and accounting for the coupling between lenses was studied in detail in \cite{Barthelemy20b} and in appendix~F1 of \cite{Barthelemy21}. These terms will tend to Gaussianise the convergence field since they characterise the introduction of random deflections along the light path which will in turn tend to diminish the impact of the non-linear clustering of matter. The heuristic picture one could form is that of clustered chunks of matter blurred by these lensing terms. It was shown that the effect matters for higher source redshifts (e.g.~CMB lensing) but is totally negligible for cosmic shear experiments.

Reduced shear corrections which at first order account to replacing the shear field by $\gamma \rightarrow \gamma + \gamma\kappa$ was studied in the LDT context in \cite{paolo} and in appendix~F2 of \cite{Barthelemy21} and was shown to induce only a percent-level change in the skewness at our scales and source-redshifts of interest. 
This tiny effect can be ignored for the analysis we propose here since the gravity-induced skewness directly implemented in the LDT formalism is itself not accurate to the percent-level because of the mixing of scales discussed in section~\ref{sec::test}. 

Finally, individual galaxies can be (de)magnified through lensing effects and thus their fluxes are de-/increased. At the flux limit of a survey, this can cause fainter sources to be included in the observed sample while they would, in the absence of lensing, be excluded. At the same time, the density of galaxies in the small region around this source appears (increased) reduced since the area of the region is also (de)magnified. As such, the net effect depends on the slope of the intrinsic, unlensed, galaxy luminosity function at the survey's flux limit. This is known as the magnification bias and it also induces a correction on the overall measured skewness of the convergence field. However, as for the reduced shear correction, it was shown in appendix~F3 of \cite{Barthelemy21} to matter very little and we hence could discard its implementation in the theoretical formalism used to analyse DES Y3-like survey data in our upcoming works.

\subsection{Photometric redshift uncertainties}

In cosmic shear surveys the redshifts of source galaxies are determined using photometric methods. Any systematic error in the photometric estimates of the galaxies can lead to biases in the redshift distribution of the source galaxies which can in turn lead to biased cosmological parameter inference. In order to include the effect of such a systematic uncertainty on the source redshift distribution, one can propagate this through the theory by updating the weak lensing kernel equation~\eqref{eq:lensing_kernel}. We could either compute this for a set of redshift distributions (obtained through different methods, for example the {\sc hyperrank} software developed in the context of DES \cite{2022MNRAS.511.2170C}) or parametrise it through a single shift parameter $\Delta z$ via \citep{gattiDES, 2022PhRvD.105b3515S,2022MNRAS.511.2170C}
\begin{equation}
    n(z) = n'(z + \Delta z) \ ,
    \label{def-photoz_error}
\end{equation}
where $n(z)$ denotes the shifted redshift distribution of the source galaxies and $n'(z)$ the default redshift distribution estimate. This simple parametrization of a single mean redshift uncertainty (one for each tomographic redshift bin) is reported to be sufficient within the statistical power of surveys such as DES Y3 2-point main analysis (see figure~10 of \cite{2022PhRvD.105b3514A} and figure~12 of \cite{2022MNRAS.511.2170C}), and will be considered sufficient for this series of papers, in coherence with the previous high-order statistics analysis of current data sets \cite{gattiDES,kids-map3}.

\section{Discussion and conclusion}
\label{sec::conclusion}

The weak lensing convergence $\kappa$ denotes a weighted projection of the three-dimensional matter density fluctuations. Intuitively, it quantifies a projected `mass' of all the late-time, non-Gaussian distributed foreground structures which contribute to the deformation of a light beam emanating from a background source galaxy on its way to us. Therefore, studying the full 1-point Probability Distribution Function (PDF) of the smoothed $\kappa$ field inside apertures (see \cite{2020PhRvD.102l3506M, 2022arXiv221210351B}) is a promising way to access the non-Gaussian cosmological information of the foreground matter density field beyond the variance (or the widely used 2PCF) and holds the potential to tighten constraints on cosmological parameters.

Unfortunately, the $\kappa$ field itself is not a direct observable as what is actually seen in lensing surveys is the \textit{cosmic shear} field --- the weak coherent distortions imprinted in the shapes of the source galaxies. Nevertheless, one can apply the widely known Kaiser-Squires (KS) inversion technique on the shear field to reconstruct the "true" inaccessible convergence field. However, the KS inversion is exact and recovers the "true" convergence only when one has information of the shear field over the entire celestial sphere\footnote{The KS inversion to reconstruct $\kappa$ at a given location involves the convolution of the shear field over the full-sky with a specific kernel.}. This is of course not the case in practice as one has access only to the shear information inside a given footprint on the sky due to survey masks (consisting of holes, complicated survey geometry and boundaries). Applying the full-sky KS inversion on the masked shear field thus results in a reconstructed $\kappa$ field which differs from the "true" inaccessible $\kappa$ field. Hence, for any $\kappa$-statistic which one desires to measure and analyse using a KS reconstructed $\kappa$ map from observed masked shear data, it is of paramount importance to correctly quantify and account for the effect of the survey mask in the theoretical modelling of the corresponding $\kappa$-statistic. In this paper we have presented for the first time how to do that in a from-first-principles theoretical modelling of the KS reconstructed $\kappa$-PDF. Our main achievements on this front can be summarised as follows:

\begin{itemize}
    \item Our reconstructed full-sky convergence PDF is obtained from the "true" one \cite{Barthelemy20a,Barthelemy21} inside the survey footprint and purely geometric factors which take into account the effect of the survey window and the survey area on the variance \eqref{masked_variance} and hence the series of cumulants \eqref{modif_phi}.
    Explicitly, this is achieved by both an accurate parametrisation (without free parameters) of the reconstructed full-sky convergence PDF as a function of the one inside the survey footprint and modifying the scaling relations of the matter density contrast field cumulants (which are needed to compute the line-of-sight projected $\kappa$ cumulants that are in turn required in the modelling of the KS reconstructed $\kappa$-PDF) by purely geometric factors which take into account the effect of the survey window and the survey area.
    \item We have applied the recipe for the full-sky KS $\kappa$-map reconstruction under the presence of a realistic survey mask (in our case, DES Year 3 survey mask) to the simulated cosmic shear data from the Takahashi suite of weak lensing N-body simulations and measured the reconstructed $\kappa$-PDF. We have tested our theoretical modelling of the same against the measurements and found excellent agreement between them within cosmic variance (figure \ref{DES_zs1}). We further find that using the baseline theoretical model for the "true" $\kappa$-PDF without accounting for the masking effect significantly deviates from simulation measurements of the reconstructed PDF. These conclusions are valid for scales and source redshifts relevant for the baseline theoretical model which is accurate on quasi-linear scales as we have also demonstrated in figure~\ref{DES_null} where we have applied a \textit{nulling strategy} to construct lensing observables less-sensitive to very small non-linear scales normally probed through scale-mixing in the projection along the line-of-sight.
    \item In preparation for an upcoming real data analysis of the KS reconstructed $\kappa$-PDF we have also discussed and laid down the strategy to include several effects such as astrophysical and measurement systematics as well as higher-order lensing corrections to the theoretical model for the reconstructed $\kappa$-PDF.
    We included a modelling for intrinsic alignments based on an adaptation of the weak lensing kernel \eqref{eq:correction_IA} and tested in simulations that baryonic feedback can be included through a rescaling of the variance \eqref{eq:PDF_baryons}. We describe how the lensing PDF is affected by a multiplicative shear biases \eqref{eq:PDF_multbias} and how to propagate photometric redshift uncertainties through our theoretical model.
\end{itemize}

Thus, our work not only presents a proper theory-based modelling framework for a real analysis of the KS reconstructed $\kappa$-PDF under the presence of realistic survey masks but it also underlines the susceptible errors when analysing any statistic from a KS reconstructed $\kappa$-map with a theoretical model that does not correctly include the E/B mode mixing due to the presence of survey masks. 

Overall, our results indicate that the $\kappa$-PDF measured from the straightforward to implement spherical-sky KS reconstructed $\kappa$-map on the observed shear field (in the presence of masks) can be treated accurately within a theoretical framework without the need for any forward-modelling simulation-based approach. In particular, though some of the systematics modelling presented in this paper might need to be improved for Stage-IV surveys, the modelling of masks in combination with LDT in the context of the BNT transform will remain valid. This paves the way for us to explore the power of the $\kappa$-PDF in probing higher-order information in current lensing surveys such as DES, and in the future using Euclid and Vera Rubin's LSST data. We will perform cosmological inference analyses on simulated and real data in the following papers of this `{\sc Making the leap}' series.

\section*{Acknowledgements}

AB's work is supported by the ORIGINS excellence cluster. CU is supported by the STFC Astronomy Theory Consolidated Grant ST/W001020/1 from UK Research \& Innovation. The authors are grateful to Ryuichi Takahashi for making the weak lensing simulation suite used in this paper publicly available. 
The authors thank Ian McCarthy for making the BAHAMAS simulation maps publicly available, as well as Tiago Castro and Klaus Dolag for providing access to convergence maps from the Magneticum simulations. The authors are indebted to Nick Kaiser for his numerous and wide-ranging contributions to the field of large-scale structures, one of which is at the centre of the present work. May he rest in peace in the knowledge that his legacy is immortal.

\section*{Data Availability}

We will release all codes written for our theoretical predictions and map-making at the release of paper II of this `{\sc Making the leap}' series. This will come with an emulator for our theoretical $\kappa_E$ PDFs which renders easier MCMC analysis of either simulated or real data.

\bibliographystyle{JHEP}
\bibliography{references}

\appendix

\section{Mode-coupling matrices in the pseudo-$C_\ell$ formalism}
\label{appendix::masking}

For completeness and better readability of this paper, we reproduce here key-results and broad derivation steps of the pseudo-$C_\ell$ formalism \citep{Hikage2011} that aims at taking masks into account in the computation of the shear power spectra. For the shear field which formally admits both $E$ and $B$ modes, we can define its full power spectra as a vector $\textbf{C}_l \equiv (C_l^{EE},C_l^{EB},C_l^{BB})$ with 
\begin{equation}
    C_l^{ij} = \frac{1}{2l+1}\sum_m \gamma_{i,lm} \gamma^*_{j,lm}, \ \text{with } i,j \in \{E,B\}.
    \label{Cl_def}
\end{equation}

We will then see that its masked power spectra $\hat{\textbf{C}}_l$ can be seen as a convolution with a mode-mixing matrix $\textbf{M}_{ll'}$ through 
\begin{equation}
    \hat{\textbf{C}}_l = \sum_{l'} \textbf{M}_{ll'} \textbf{C}_{l'}.
\end{equation}
The formalism thus takes into account transfers of power due to masking from pure E-modes spectra to pure B-modes and all other possible combinations. In this work, since we neglect any source of shear B-modes prior to masking, we are mostly interested in the element $M_{l l^{\prime}}^{EE, EE}$ that allows to compute the contribution from the unmasked shear $EE$ spectrum to the masked $\hat{C}_l^{EE}$. 
Let us now sketch the derivation of these terms.

In the presence of a footprint $K(\theta,\phi)$ (in our case the DES Y3 mask) the observed shear field becomes
\begin{equation}
    \hat{\gamma}_1(\theta, \phi)+i \hat{\gamma}_2(\theta, \phi)=K(\theta, \phi)\left(\gamma_1(\theta, \phi)+i\gamma_2(\theta, \phi)\right).
\end{equation}
Written in terms of spherical harmonics coefficients, this leads to the generation of both pseudo-E and pseudo-B modes
\begin{equation}            
\hat{\gamma}_{E, l m} \pm i \hat{\gamma}_{B, l m}=\int {\rm d} \Omega\left[K(\theta, \phi)\left(\gamma_1(\theta, \phi)\pm i\gamma_2(\theta, \phi)\right)\right] { }_{ \pm 2}Y_{l m}^*(\theta, \phi)
\end{equation}
which can be expressed in terms of the original E and B modes via
\begin{equation}
    \hat{\gamma}_{E, l m} \pm i \hat{\gamma}_{B, l m}=\sum_{l^{\prime} m^{\prime}}\left(\gamma_{E, l m} \pm i \gamma_{B, l m}\right) { }_{ \pm 2}W_{l l^{\prime} m m^{\prime}}
    \label{A5}
\end{equation}
where
\begin{multline}
 { }_{ \pm 2} W_{l l^{\prime} m m^{\prime}}=\int {\rm d} \Omega { }_{ \pm 2} Y_{l^{\prime} m^{\prime}}(\theta, \phi) K(\theta, \phi) { }_{ \pm 2} Y_{l m}^*(\theta, \phi) \\ =
 \sum_{l^{\prime \prime} m^{\prime \prime}} K_{l^{\prime \prime} m^{\prime \prime}}(-1)^m \sqrt{\frac{(2 l+1)\left(2 l^{\prime}+1\right)\left(2 l^{\prime \prime}+1\right)}{4 \pi}}
 \left(\begin{array}{ccc}
l & l^{\prime} & l^{\prime \prime} \\
\pm 2 & \mp 2 & 0
\end{array}\right)\left(\begin{array}{ccc}
l & l^{\prime} & l^{\prime \prime} \\
m & m^{\prime} & m^{\prime \prime}
\end{array}\right).
\end{multline}
In the above, $K_{lm}$ are the conjugate spherical harmonics coefficients of the window function $\int {\rm d} \Omega K(\theta,\phi) Y^*_{lm}(\theta,\phi)$, and $\left(\begin{array}{ccc} l & l^{\prime} & l^{\prime \prime} \\ m & m^{\prime} & m^{\prime \prime}\end{array}\right)$ are the usual Wigner $3j$ symbols. Finally, combining the last result in equation \eqref{A5} with the definition of the power spectra~\eqref{Cl_def} for the pseudo-E and pseudo-B modes, one can then access the mode-mixing matrix coefficients and obtain
\begin{equation}
M_{l l^{\prime}}^{E E, E E}=M_{l l^{\prime}}^{B B, B B} = \frac{2 l^{\prime}+1}{8 \pi} \sum_{l^{\prime \prime}}\left(2 l^{\prime \prime}+1\right) K_{l^{\prime \prime}} \left[1+(-1)^{l+l^{\prime}+l^{\prime \prime}}\right]
\left(\begin{array}{ccc}
l & l^{\prime} & l^{\prime \prime} \\
2 & -2 & 0
\end{array}\right)^2,
\end{equation}

\begin{equation}
M_{l l^{\prime}}^{E E, BB}=M_{l l^{\prime}}^{B B, EE} = \frac{2 l^{\prime}+1}{8 \pi} \sum_{l^{\prime \prime}}\left(2 l^{\prime \prime}+1\right) K_{l^{\prime \prime}} \left[1-(-1)^{l+l^{\prime}+l^{\prime \prime}}\right]
\left(\begin{array}{ccc}
l & l^{\prime} & l^{\prime \prime} \\
2 & -2 & 0
\end{array}\right)^2,
\end{equation}

\begin{equation}
    M_{l l^{\prime}}^{E B, E B}=\frac{2 l^{\prime}+1}{4 \pi} \sum_{l^{\prime \prime}}\left(2 l^{\prime \prime}+1\right) K_{l^{\prime \prime}}\left(\begin{array}{ccc}
l & l^{\prime} & l^{\prime \prime} \\
2 & -2 & 0
\end{array}\right)^2,
\end{equation}

with $K_l \equiv 1/(2l+1) \sum_m K_{lm} K_{lm}^*$, the power spectrum of the mask. All other elements of the mode-mixing matrix are equal to zero.

\section{Masking effects on the Kaiser-Squires reconstructed convergence moments}
\label{appendix::KS-moments}

Here we quantify the effect of the KS reconstruction scheme on the moments/cumulants of the $\kappa_E$ field. The goal of this appendix is i) to illustrate that the KS reconstruction has an important quantifiable effect on the variance and skewness, thus justifying the need for its theoretical modelling and ii) to demonstrate that the theoretical model captures well this reconstruction effect on the variance and the skewness. 

We provide in table~\ref{table:cumulants} the values of the variance and skewness of the true convergence fields as measured in the Takahashi simulation (i.e.~without any reconstruction) and the analytically computed (theory) values with a \verb|halofit| power spectrum \cite{2012ApJ...761..152T}. We also provide those same values for the reconstructed $\kappa_E$ field. As expected from the mode-mixing formalism used in this paper, the modelling of the KS reconstruction perfectly recovers the effect on the variance as measured. The roughly $\sim 15\%$ loss in the value of the skewness as measured in the simulations (cf.~the fractional difference of the measured $\langle\kappa^3\rangle_{c, \rm true}$ and $\langle\kappa_E^3\rangle_{c, \rm unmasked}$ values) is also well recovered by the theory modelling (cf.~the fractional difference of the theory $\langle\kappa^3\rangle_{c, \rm true}$ and $\langle\kappa_E^3\rangle_{c, \rm unmasked}$ values). As expected, the reconstruction does not significantly improve nor reduce the accuracy of the large-deviations formalism describing the underlying dynamics of the matter density field. The values given in table~\ref{table:cumulants} are supplementary to figure~\ref{DES_zs1} and are thus taken from a reconstructed field under the DES Y3 mask and for sources in the Takahashi simulation mimicking the DES Y3 4$^{\rm th}$ bin displayed in figure~\ref{n_z}.
\begin{table}[ht!]
\hspace{-1.5cm}
\begin{tabular}{|c||ccc|cc|}
\hline
       & $\sigma^2_{\kappa, {\rm true} }$ ($10^{-5}$) & $\sigma^2_{\kappa_E, {\rm FS}}$ ($10^{-6}$) & $\sigma^2_{\kappa_E, {\rm unmasked}}$ ($10^{-5}$) & $\langle\kappa^3\rangle_{c, \rm true}$ ($10^{-8}$) & $\langle\kappa_E^3\rangle_{c, \rm unmasked}$ ($10^{-8}$)\\ 
\hline
 Measured & $2.29 \pm 0.02$ & $2.43 \pm 0.05$ & $2.07 \pm 0.05$ & $5.56 \pm 0.18$ & $4.86 \pm 0.47$\\  
 Theory & $2.30$ & $2.44$ & $2.08$ & $4.97$ & $4.05$\\
\hline
\end{tabular}
\caption{Smoothed second and third cumulants ($\theta = 20$ arcmin) of the reconstructed $\kappa_E$ under the DES Y3 mask on the shear field and for sources in the 4$^{\rm th}$ DES Y3 redshift bin. All measurements are made in the Takahashi simulation and this table is supplementary to figure~\ref{DES_zs1}. We give values for the reconstructed full-sky $\kappa_E$ variance and the variance and skewness values across the pixels under the mask that we take into account as described at the end of section~\ref{sec::reconstruct_PDF}.}
\label{table:cumulants}
\end{table}

\section{Square mask for sources in the DES Y3 4$^{\rm th}$ bin}
\label{sec::SQUAREmask}

We here replicate the results of section~\ref{sec::DESmask}, that is a source distribution (shown in figure~\ref{n_z}) mimicking the 4$^{\rm th}$ source bin of the DES Y3 analysis but this time using a square mask without holes of roughly the same area as the DES Y3 mask. Since by construction the KS effects on the variance are perfectly taken into account, any discrepancy in the modelling of the KS effects comes at the level of the skewness of the reconstructed $\kappa_E$ field. As before, we quantify the quality of the KS modelling by comparing the residuals of the theoretical KS PDF to the simulated KS reconstruction PDF and the residuals of those same PDFs without any reconstruction. That way we visually do not include the intrinsic quality of the LDT prediction in our assessment. 

\begin{figure}
    \hspace{-1.1cm}
    \begin{subfigure}{0.49\columnwidth}
        \hspace{-0.9cm}
        \includegraphics[width=1.25\columnwidth]{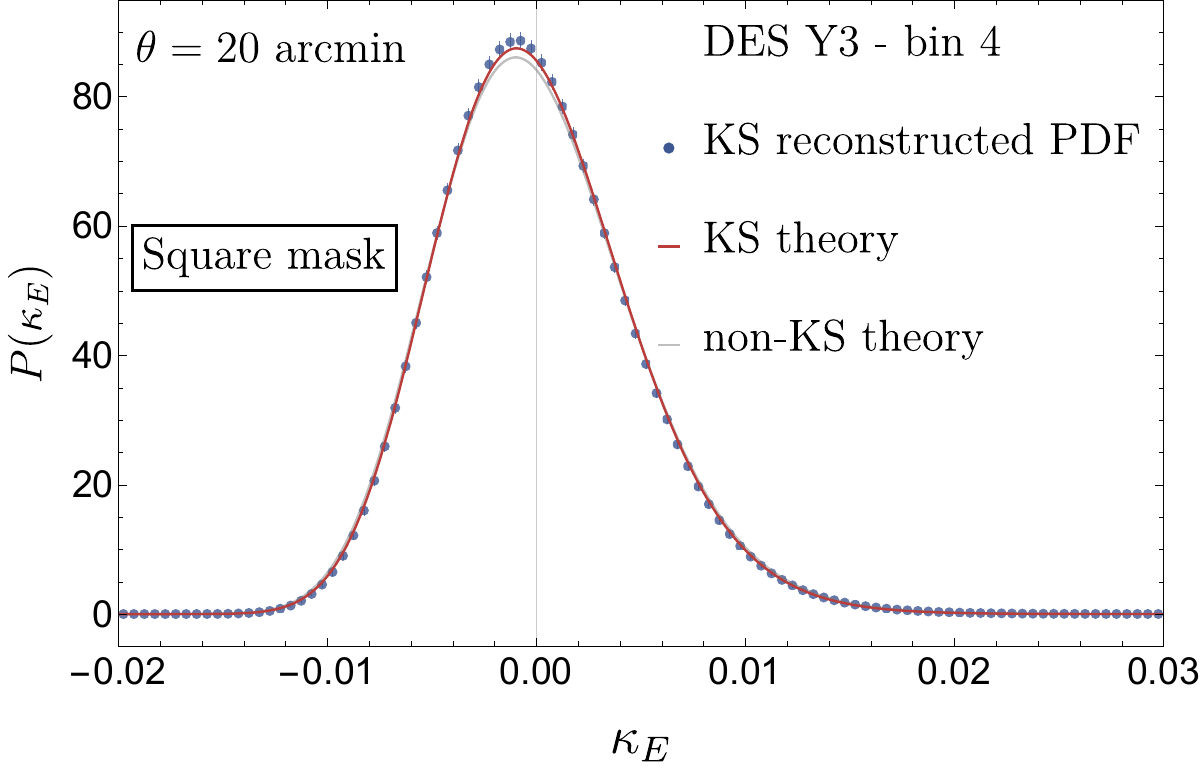}
    \end{subfigure}
    \hfill
    \begin{subfigure}{0.49\columnwidth}
        \hspace{-0.3cm}
        \includegraphics[width=1.28\columnwidth]{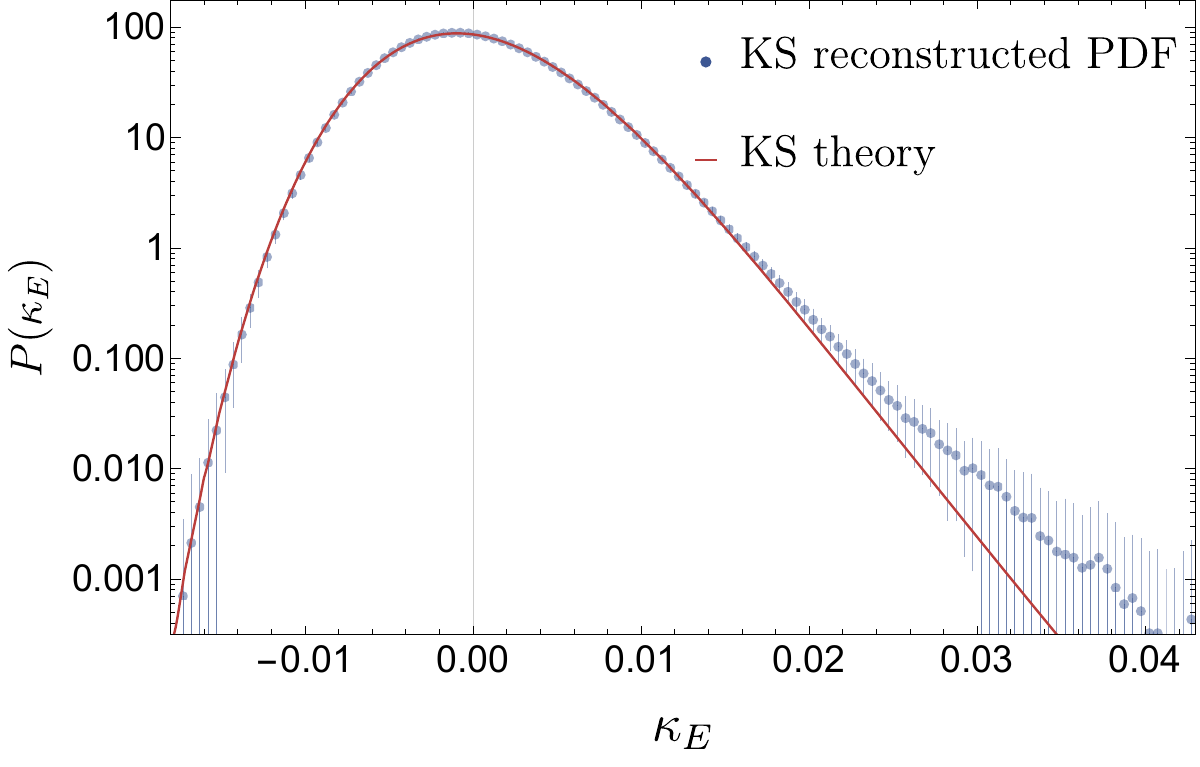} 
    \end{subfigure}
    \begin{subfigure}{\columnwidth}
    \centering
        \includegraphics[width=1.05\columnwidth]{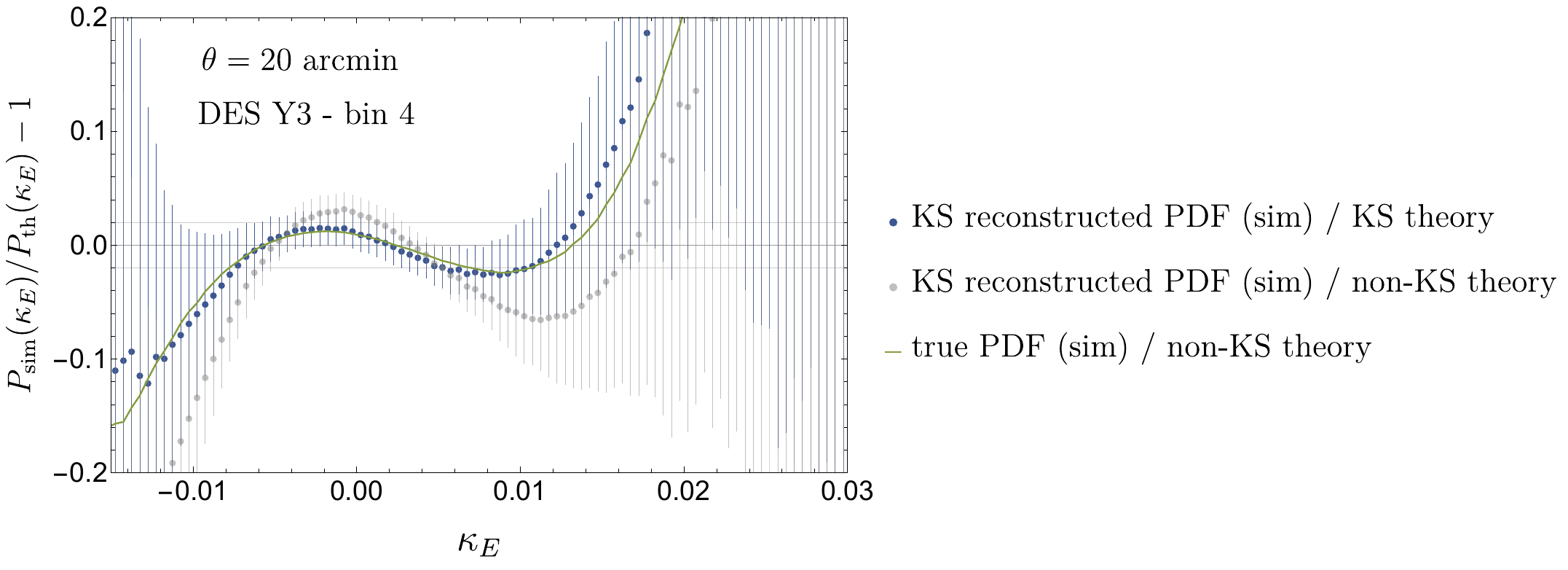} 
    \end{subfigure}
    \caption{\textit{Top panels:} PDF of the Kaiser-Squires weak-lensing convergence map for the DES Y3-like 4$^{\rm th}$ bin in the Takahashi simulation and under an idealised square-patch mask of $70 \times 70$ deg$^2$. The red line is the theory described by equation~\eqref{modif_phi} and the blue points with error bars are the mean and $1\sigma$ fluctuations of the measured PDF in 108 Takahashi maps. The grey solid line is the theory described by equation~\eqref{projection} that is without taking into account the KS reconstruction. \textit{Bottom panel:} Residuals between the theory and the simulation. The green line is the residual of the original non-KS theory to the true, non-reconstructed simulated convergence PDF. It is highly compatible with the residuals of the KS-theory to the simulated reconstructed KS PDF (blue points). Though their inclusion allows to recover the full predictive power of our LDT formalism, neglecting the KS reconstruction effects is less imperative for this simple idealised square-mask, as opposed to the more realistic DES Y3 mask. This is understandable as discussed in appendix~\ref{sec::SQUAREmask}.}
    \label{square_zs1}
\end{figure}

Looking at the bottom panel of figure~\ref{square_zs1} we observe that the green line matches better the mean residuals of the theoretical KS PDF to the simulated KS reconstruction PDF (blue points) than what we could observe in the case of the true DES Y3 mask in figure~\ref{DES_zs1}. At the same time the impact of neglecting the KS effects in the modelling are less pronounced for this idealised mask (grey points). This is understandable to a certain extent. Indeed, the presence of multiple holes of various sizes within the DES Y3 mask makes the mixing of wave-modes in the E to E/B modes leakage more complicated to a degree where our mitigation (performed by restricting ourselves to the less affected pixels, cf.~section~\ref{sec::reconstruct_PDF}) is less effective than in this simpler square-mask case. Note that similarly, finding it harder to mitigate the E/B mode leakage, would happen if the area of the square patch is small since the boundary effects would then affect more seriously the entire available area under the mask. The next step to improve the modelling would be to compute explicitly the convolution kernel on the shear bispectrum induced by the mask and include the new correction on the skewness in a manner analogous to how we would include higher-order weak lensing correction, i.e.~by modification of the spherical collapse parameter $\nu$ in equation~\eqref{collapse} \citep{Barthelemy20b}. This however does not seem to be necessary at this stage for a DES Y3 analysis.

\end{document}